\documentclass[12pt]{iopart}
\usepackage{iopams}
\usepackage{graphicx}
\usepackage{amssymb}
  
\begin{document}

\title[In-degree and out-degree distributions of directed node duplication networks]
{
Analytical results for the in-degree and out-degree distributions
of directed random networks that grow by node duplication
}

\author{Chanania Steinbock, Ofer Biham \& Eytan Katzav}
\address{
Racah Institute of Physics, 
The Hebrew University, 
Jerusalem 91904, Israel}
\eads{\mailto{chansteinbock@gmail.com}, 
\mailto{biham@phys.huji.ac.il}, 
\mailto{eytan.katzav@mail.huji.ac.il}}

\begin{abstract}

We present exact analytical results for the degree distribution 
in a directed network model that grows by node duplication.
Such models are useful in the study of the structure and growth dynamics of
gene regulatory networks and scientific citation networks.
Starting from an initial seed network, at each time step
a random node, referred to as a mother node, is selected for duplication.
Its daughter node is added to the network 
and duplicates each outgoing link of the mother node
with probability $p$.
In addition, the daughter node forms a directed link to the 
mother node itself. 
Thus, the model is referred to as the corded directed-node-duplication
(DND) model.
The corresponding undirected node duplication model was
studied before and was found to exhibit a power-law degree distribution.
We obtain analytical results for the in-degree distribution 
$P_t(K_{\rm in}=k)$, 
and for the
out-degree distribution 
$P_t(K_{\rm out}=k)$,
of the corded DND network at time $t$. 
It is found that the in-degrees follow a shifted power-law distribution,
so the network is asymptotically scale free.
In contrast, the out-degree distribution is a narrow distribution, 
that converges to a Poisson distribution in the limit of $p \ll 1$
and to a Gaussian distribution in the limit of $p \simeq 1$.
Such distinction between a broad in-degree distribution and a 
narrow out-degree distribution is common in empirical networks
such as scientific citation networks.
Using these distributions we calculate the mean degree 
$\langle K_{\rm in} \rangle_t = \langle K_{\rm out} \rangle_t$,
which converges to $1/(1-p)$ in the large network limit, for the
whole range of $0 < p < 1$.
This is in contrast to the corresponding undirected network,
which exhibits a phase transition at $p=1/2$ such that for
$p>1/2$ the mean degree diverges in the large network limit.
We also present analytical results for the distribution
of the number of upstream nodes, 
$P_t(N_{\rm up}=n)$,
and for the distribution of the number of downstream nodes,
$P_t(N_{\rm down}=n)$, from a random node.
We show that the mean values
$\langle N_{\rm up} \rangle_t = \langle N_{\rm down} \rangle_t$
scale logarithmically with the network size.
This means that in the large network limit only a diminishing fraction
of pairs of nodes are connected by directed paths,
unlike the corded undirected node duplication network that consists of
a single connected component.
Therefore, the corded DND network is not a small-world network.

\end{abstract}

\section{Introduction}

The increasing interest in the field of 
complex networks in recent years is motivated by 
the realization that a large variety of systems and processes
in physics, chemistry, biology, engineering, and society 
can be usefully described by network models
\cite{Albert2002,Caldarelli2007,Havlin2010,Newman2010,Estrada2011b,Barrat2012}.
These models consist of nodes and edges, where the nodes
represent physical objects, while the edges represent the
interactions between them.
Many of these networks are scale-free,
which means that they exhibit power-law degree distributions
\cite{Barabasi1999,Jeong2000,Krapivsky2000,Krapivsky2001,Vazquez2003}.
The most highly connected nodes, called hubs, 
play a dominant role in dynamical processes on these networks.
A common feature of complex networks is the small-world property,
namely the fact that the mean distance and the diameter 
scale like $\ln N$,
where $N$ is the network size
\cite{Milgram1967,Watts1998,Chung2002,Chung2003}.
Moreover, it was shown that undirected scale-free networks are generically
ultrasmall, namely their mean distance and diameter scale like
$\ln \ln N$
\cite{Cohen2003}.

To gain insight into the structure of complex networks, 
it is useful to study the growth dynamics that gives rise to these structures.
In general, it appears that many of the networks encountered in biological,
ecological and social systems grow step by step, by the addition of new nodes
and their attachment to existing nodes. 
A common feature of these growth processes is
the preferential attachment mechanism, in which the likelihood of
an existing node to gain a link to the new node 
is proportional to its degree.
It was shown that growth models based on preferential
attachment give rise to scale-free networks,
which exhibit power-law degree distributions
\cite{Albert2002,Barabasi1999}. 
The effect of node duplication (ND) processes 
on network structure 
was studied using an undirected network growth model
in which at each time step a random node, 
referred to as a mother node,
is selected for duplication
and its daughter node duplicates
each link of the mother node with probability $p$
\cite{Bhan2002,Kim2002,Chung2003b,Krapivsky2005,Ispolatov2005,Ispolatov2005b,Bebek2006,Li2013}.
In this model the daughter node does not form a link
to the mother node, and thus in the following it
is referred to as the
uncorded ND model. 
It was shown that for $0 < p < 1/2$ 
the resulting network exhibits a power law
degree distribution of the form 

\begin{equation}
P(K=k) \sim k^{-\gamma}.
\label{eq:Pkkg}
\end{equation}

\noindent
For 
$0 < p < 1/e$,
where $e$ is the base of the natural logarithm,
the exponent is given by the nontrivial solution of
the equation
$\gamma = 3 - p^{\gamma-2}$, 
while for 
$1/e \le p < 1/2$ 
it takes the value 
$\gamma=2$
\cite{Ispolatov2005}.
For $1/2 \le p \le 1$ the degree distribution does not converge 
to an asymptotic form.

Recently, a different 
variant of an undirected
node duplication model
was introduced and studied
\cite{Lambiotte2016,Bhat2016,Steinbock2017}.
In this model, referred to as the corded ND model, 
at each time step a random mother node, M,
is selected for duplication.
The daughter node, D, is added to the network.
It forms an undirected link to its mother node, M,
and is also connected with probability $p$ to each 
neighbor of M.
It was shown that for 
$0 < p < 1/2$ 
the corded ND model generates a sparse network,
while for
$1/2 \le p \le 1$ the model gives rise to a dense network in 
which the mean degree increases with the network size
\cite{Lambiotte2016,Bhat2016}.
For $0 < p < 1/2$ 
the degree distribution 
of this network follows 
a power-law distribution,
given by Eq. (\ref{eq:Pkkg}),
where the exponent 
$\gamma$
is given by the non-trivial solution of the
equation 
$\gamma = 1 + p^{-1} - p^{\gamma-2}$
\cite{Lambiotte2016,Bhat2016}.
In the limit of $p \rightarrow 0$, the exponent 
diverges like
$\gamma \sim 1/p$.
This model is suitable for the description of acquaintance networks,
in which a newcomer who has a friend in the 
community becomes acquainted with other members 
\cite{Toivonen2009}.
Unlike the uncorded ND model, 
the formation of triadic
closures is built-in to the dynamics of the corded ND model. 
This means that once the
daughter node forms a link to a neighbor of the mother node, it completes
a triangle in which the mother, neighbor and daughter nodes are all connected 
to each other. The formation of triadic closures is an essential property of
the dynamics of social networks
\cite{Granovetter1973}. 
The formation of triadic closures is in sharp contrast to
configuration model networks, which exhibit a local
tree-like structure.
Interestingly, many empirical networks exhibit a high 
abundance of triangles,
both in undirected networks
\cite{Newman2001b}
and in directed networks,
where most triangles form feed-forward loops (FFLs), while triangular
feedback loops are rare
\cite{Milo2002,Alon2006}.
Unlike configuration model networks
\cite{Molloy1995,Molloy1998,Newman2001}, 
which may include small, isolated components,
the corded ND network consists of a single connected component.
Therefore, it does not exhibit a percolation transition.

In this paper we introduce a directed version of the corded node
duplication model, referred to as the
corded directed node duplication (DND) model.
In this model, at each time step a random mother node is chosen for 
duplication. The daughter node forms a directed
link to the mother node and with probability $p$ to each outgoing
neighbor of the mother node.
This model may be useful in the study of
gene regulatory networks.
These are directed networks that evolve by gene duplication
\cite{Ohno1970,Teichmann2004}.
It also describes the structure and dynamics of scientific citation networks
\cite{Redner1998,Redner2005,Radicchi2008,Golosovsky2012,Golosovsky2017a,Golosovsky2017b},
in which the nodes represent papers, while the links
represent citations.
Scientific citation networks are
directed networks, with links pointing from the later (citing) paper
to the earlier (cited) paper.
A paper A, citing an earlier paper B,
often also cites one or several papers C,
which were cited in B
\cite{Peterson2010}.
The resulting network module is a triangle, or triadic closure,
which includes three directed links, from A to B, from B to C and
from A to C and thus resembles the FFL structure.
However, the links of this module point backwards, 
and thus it may be more suitable to refer to it as a 
feed-backward loop (FBL).
Due to the directionality of the links, each node exhibits both an in-degree,
which is the number of incoming links and an out-degree which is the number
of outgoing links.
Therefore, the degree distribution consists of two separate distributions,
namely the distribution $P_t(K_{\rm in}=k)$ of in-degrees and the distribution
$P_t(K_{\rm out}=k)$ of out-degrees, at time $t$. 
These two distributions are related to each other by the constraint that their means
must be equal, namely
$\langle K_{\rm in} \rangle_t = \langle K_{\rm out} \rangle_t$.

We present exact analytical results for 
the in-degree distribution and the out-degree distribution of the
corded DND network.  
It is found that the in-degrees follow a shifted power-law distribution
while the out-degrees follow a narrow distribution  
that converges to a Poisson distribution in the limit of $p \ll 1$
and to a Gaussian distribution in the limit of $p \simeq 1$.
Since the network is directed not all pairs of nodes are connected by
directed paths, unlike the corresponding undirected network that
consists of a single connected component.
We present analytical results for the distribution
of the number of upstream nodes
$P_t(N_{\rm up}=n)$,
and for the distribution of the number of downstream nodes
$P_t(N_{\rm down}=n)$,
and show that 
$\langle N_{\rm up} \rangle = \langle N_{\rm down} \rangle$
are logarithmic in the network size.
This means that in the large network limit only a diminishing fraction
of pairs of nodes are connected by directed paths.
Therefore, the corded DND network is not a small-world network,
unlike the corresponding undirected network.

The paper is organized as follows.
In Sec. 2 we present the corded DND model.
In Sec. 3 we analyze the structure of the backbone tree
of the corded DND network, which consists of the deterministic 
links between the mother and daughter nodes.
The distribution of the number of nodes upstream of a random
node, denoted by $P_t(N_{\rm up}=n)$, is calculated in Sec. 4.
The distribution of the number of nodes downstream of a random
node, denoted by $P_t(N_{\rm down}=n)$, is calculated in Sec. 5.
Note that $P_t(N_{\rm up}=n)$ and $P_t(N_{\rm down}=n)$
are properties of the backbone tree, which do not depend on
the parameter $p$.
In Sec. 6 we consider the temporal evolution of the mean degree
$\langle K \rangle_t = \langle K_{\rm in} \rangle_t = \langle K_{\rm out} \rangle_t$.
In Sec. 7 we calculate the in-degree distribution 
$P_t(K_{\rm in}=k)$, and in Sec. 8 we calculate
the out-degree distribution $P_t(K_{\rm out}=k)$.
The results are discussed in Sec. 9 and summarized in Sec. 10.
In Appendix A we prove a mathematical identity, based on the q-Pochhammer
symbol, which is used in the analysis of the out-degree distribution.
In Appendix B we analyze the behavior of the out-degree distribution
$P_t(K_{\rm out}=k)$,
in the limit of $p \ll 1$, while in Appendix C we consider its
behavior in the limit of $p \simeq 1$.

\begin{figure}
\centerline{
\includegraphics[width=10cm]{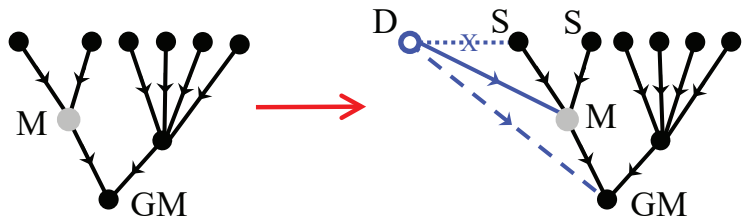}
}
\caption{
Illustration of the corded DND model.
A random node, referred to as a mother node, M (gray circle)
is selected for duplication. The newly formed daughter node, D
(empty circle) deterministically acquires a directed link (solid line) 
to the mother node. It also acquires, with probability $p$,
a directed link (dashed line) to each one of the outgoing neighbors of M.
In this example, D forms a directed link
to its grandmother node, denoted by GM.
The model does not allow D to form links to its sister nodes, denoted by S,
because they are incoming neighbors rather than outgoing neighbors of M.
}
\label{fig:1}
\end{figure}

\section{The corded directed node duplication model}

In the corded DND model,
at each time step during the growth phase of the
network, a random node,
referred to as a mother node,
is selected for duplication.
The daughter node is added to the network, 
forming a directed link to the mother node.
Also, with probability $p$, it forms a directed link
to each outgoing neighbor
of the mother node (Fig. \ref{fig:1}).
The growth process starts from an initial 
seed network of $N_0=s$ nodes.
Thus, the network size
after $t$ time steps is
 
\begin{equation}
N_t = t + s.
\end{equation}

\begin{figure}
\centerline{
\includegraphics[width=6.5cm]{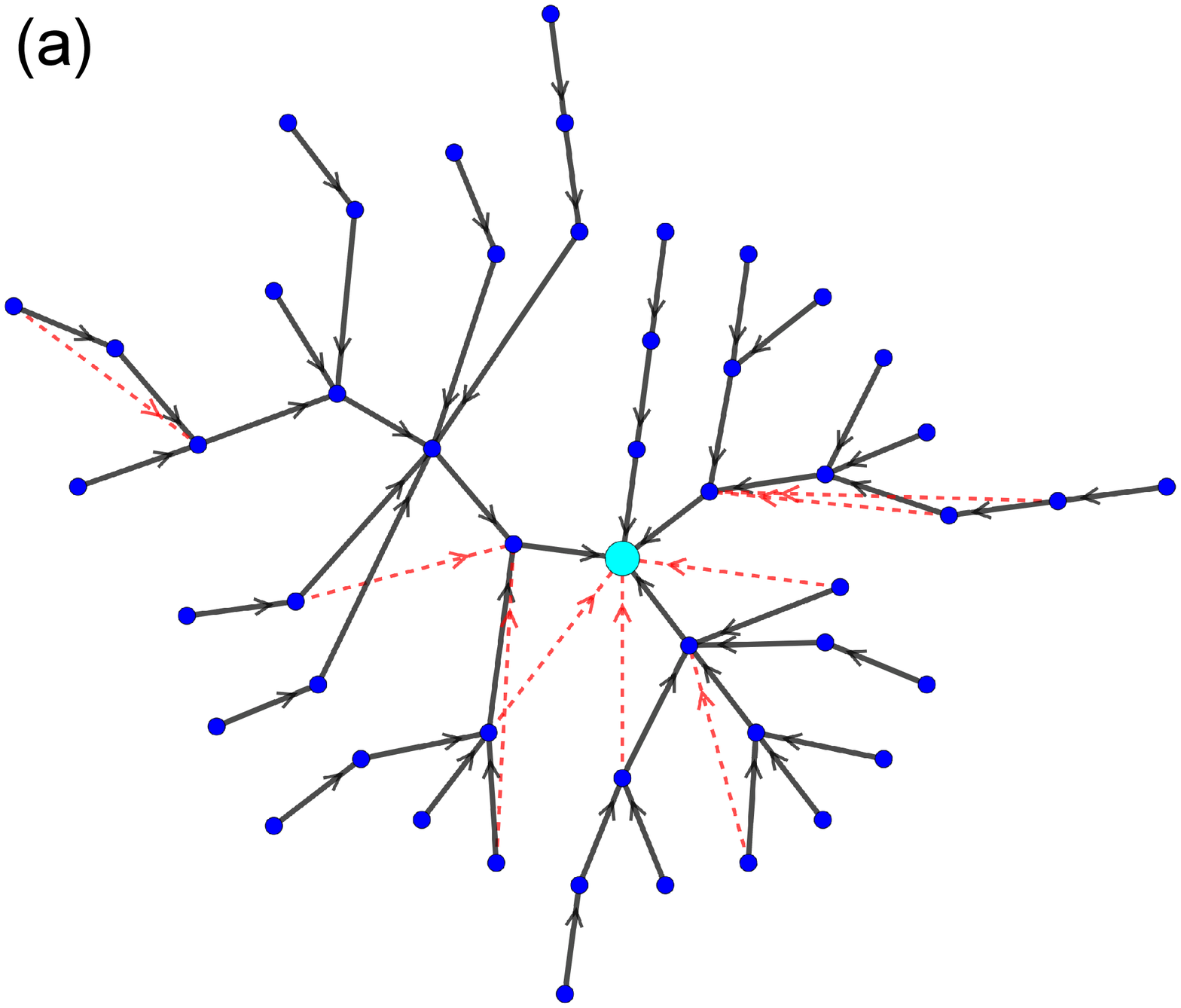} 
\includegraphics[width=6.5cm]{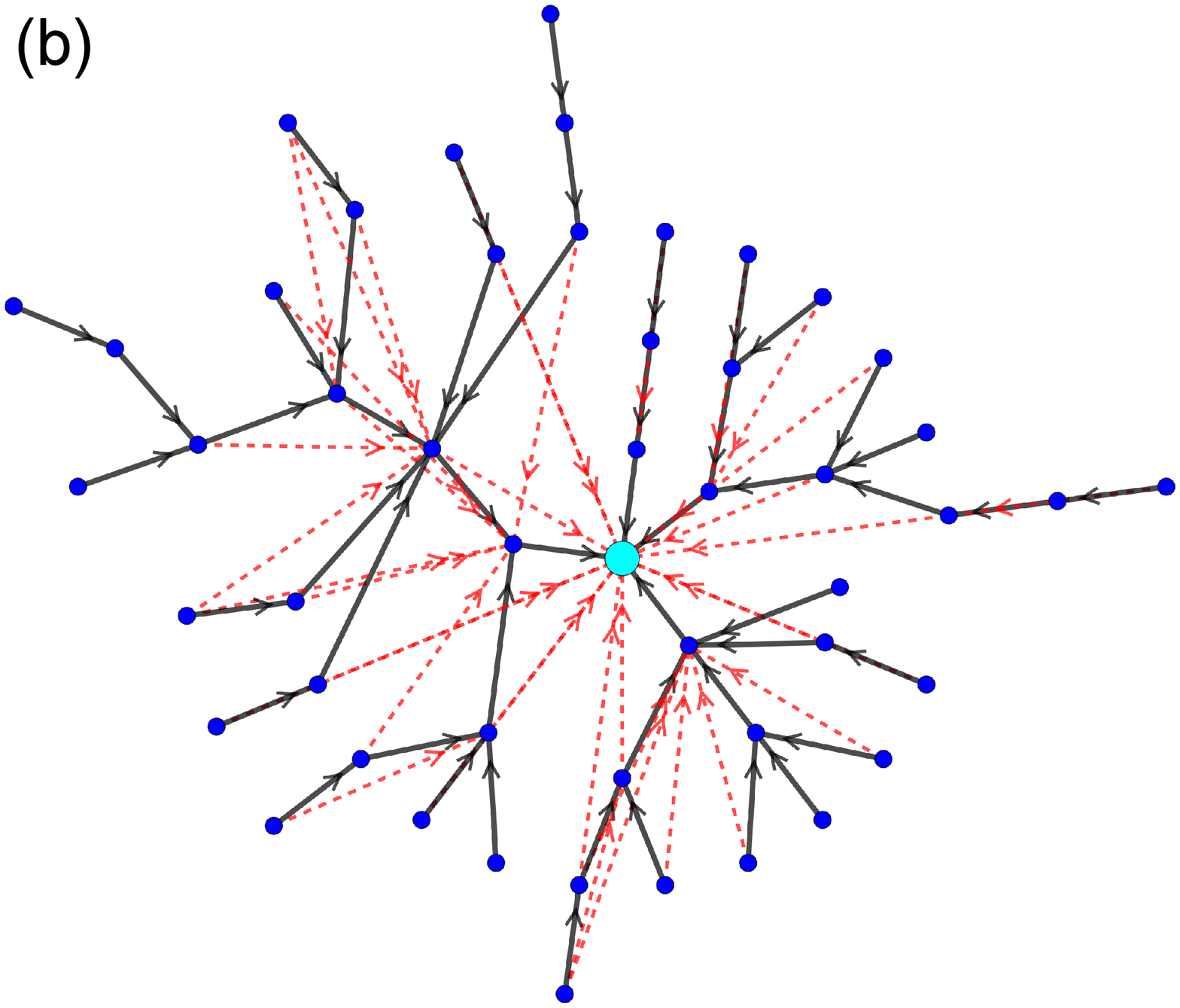}
}
\caption{
Two instances of corded DND networks of size $N=50$,
with $p=0.2$ (a) and $p=0.5$ (b).
Both networks were grown from a seed network that consists of
two nodes connected by a directed link.
For the sake of comparison, both instances are formed around
the same backbone tree (solid lines).
The sink node, which can be reached from all the other nodes in
the network via directed paths is shown by a large circle.
The probabilistic links (dashed lines) essentially decorate the
tree. Increasing $p$ makes the network denser.
}
\label{fig:2}
\end{figure}

\noindent
In Fig. \ref{fig:2}
we present two instances of the corded DND network, of size $N_t=50$,
which were formed around the same backbone tree.
Both networks were grown from a seed network of size $s=2$,
with $p=0.2$ [Fig. \ref{fig:2}(a)] and $p=0.5$ [Fig. \ref{fig:2}(b)].
Thus, each network instance includes $N_t-1=49$ deterministic links
(solid lines).
The network of Fig. \ref{fig:2}(b) is denser and includes $35$
probabilistic links (dashed lines), 
compared to $9$ probabilistic links in 
Fig. \ref{fig:2}(a).

Upon formation, the in-degree of the daughter node is zero. 
The incoming links are gradually formed as the daughter node matures.
Since the mother node at time $t$ is selected randomly from
all the $N_t$ nodes in the network, its in-degree
is effectively drawn from the in-degree distribution $P_t(K_{\rm in}=k)$.
The mother node gains an incoming link from the daughter node, thus
its in-degree increases by $1$.
The daughter node gains one outgoing link to the mother node,
and with probability $p$ it duplicates outgoing
links of the mother node.
Thus, in the case that all the outgoing links of the mother node are duplicated, 
the out-degree of the daughter node becomes
$k_{\rm out}^{\rm D} = k_{\rm out}^{\rm M} + 1$.
In the case that none of the outgoing links of the mother node
are duplicated the out-degree of the daughter node becomes
$k_{\rm out}^{\rm D}=1$.

In order to obtain a network that consists of a
single connected component, it is required
that the seed network consist of a single connected
component.
The size of the seed network is denoted by $s$. 
Since the corded DND model generates oriented networks with no bidirectional edges,
we restrict the analysis to seed networks in which a pair of nodes cannot
be connected in both directions. 
Moreover, we consider only acyclic seed networks, 
namely networks that do not include any
directed cycles.
Finally, we focus on seed networks that include a single
sink node, namely a node that has only incoming links and
no outgoing links.
The sink node can be reached via directed paths from all the nodes
in the seed network.
The in-degree distribution of the seed network is denoted by 
$P_0(K_{\rm in}=k)$
and its out-degree distribution is denoted by
$P_0(K_{\rm out}=k)$.
Clearly, the mean of the in-degree distribution
and the mean of the out-degree distribution
are equal to each other.
We thus denote
$\langle K \rangle_0 = \langle K_{\rm in} \rangle_0 = \langle K_{\rm out} \rangle_0$.

To avoid memory effects, which may slow down the
convergence to the asymptotic structure, it is often convenient to use
a seed network that consists of a single node,
namely $s=1$.
In this case
the in-degree and out-degree distributions of the seed network are given by
$P_0(K_{\rm in}=k) = P_0(K_{\rm out}=k) = \delta_{k,0}$,
where $\delta_{k,k'}$ is the Kronecker delta, which satisfies
$\delta_{k,k'}=1$ for $k=k'$ and $\delta_{k,k'}=0$ for $k \ne  k'$.
Another interesting choice for the seed network is a
linear chain of $s$ nodes, in which all the links are
in the same direction.
In this case, the 
initial degree distribution is
$P_0(K_{\rm in}=k) = P_0(K_{\rm out}=k) = (1/s) \delta_{k,0} + (1-1/s) \delta_{k,1}$.
While at early times the size and structure of the seed network has a strong
effect on the structure of the growing network, in the long time limit of
$t \gg s$ this effect vanishes.
Thus, in the long time limit the scaling behavior can be expressed in
terms of the time $t$, which represents the number of nodes that
were added to the network during the growth phase.

\section{The backbone tree}

The mother-daughter links in the
corded DND network form a random directed tree structure,
which serves as a backbone tree for the resulting network.
The backbone tree is a random directed recursive tree
\cite{Smythe1995,Drmota1997,Drmota2005}.
To study its properties, one can take the limit of $p=0$,
in which the corded DND network is reduced to the backbone tree.
As a seed network we take a directed tree network of $s$ nodes,
which includes a single sink node, whose in-degree distribution is 
$P_0(K_{\rm in}=k)$ and whose out-degree distribution is $P_0(K_{\rm out}=k)$.

The in-degree distribution of the backbone tree,
denoted by
$P_t(K_{\rm in}=k)$,
evolves in time according to

\begin{equation}
P_{t+1}(K_{\rm in}=k) = 
\frac{
 (N_t-1) P_t(K_{\rm in}=k) 
+ P_t(K_{\rm in}=k-1) 
+ \delta_{k,0} }
{N_t+1},
\label{eq:recK}
\end{equation}

\noindent
where the initial condition is given by the in-degree distribution,
$P_0(K_{\rm in}=k)$,
of the seed network.
The first term on the right hand side 
of Eq. (\ref{eq:recK})
accounts for the existing nodes in the network
(apart from the mother node),
whose in-degrees are not affected by the node duplication step.
The second term accounts for the in-degree of the
mother node, which increases by $1$ due to the incoming link from the daughter node.
The third term accouts for the in-degree of D,
which is $k=0$.
Subtracting $P_t(K_{\rm in}=k)$ from both sides 
of Eq. (\ref{eq:recK})
and replacing the difference on the left
hand side by a time derivative, we obtain

\begin{equation}
\frac{\partial}{\partial t} P_t(K_{\rm in}=k) = 
- \frac{   2 P_t(K_{\rm in}=k) 
- P_t(K_{\rm in}=k-1)
- \delta_{k,0} }{t+s+1}.
\label{eq:mera2}
\end{equation}

\noindent
Using a generating function formulation, we rewrite Eq. (\ref{eq:mera2})
in the form

\begin{equation}
\frac{\partial}{\partial t} G_t^{\rm in}(x) = - \frac{ (2-x) G_t^{\rm in}(x) - 1 }{t+s+1},
\label{eq:Ginb}
\end{equation}

\noindent
where

\begin{equation}
G_t^{\rm in}(x) = \sum_{k=0}^{\infty} x^k P_t(K_{\rm in}=k),
\end{equation}

\noindent
is the generating function of the in-degree distribution at time $t$
and $0 \le x \le 1$.
Solving Eq. (\ref{eq:Ginb}), we obtain

\begin{equation}
G_t^{\rm in}(x) = \left[ G_0^{\rm in}(x) - \frac{1}{2-x} \right] 
\frac{1}{t_s^{2-x} } + \frac{1}{2-x},
\label{eq:Gtbbt}
\end{equation}

\noindent
where

\begin{equation}
G_0^{\rm in}(x) = \sum_{k=0}^{\infty} x^k P_0(K_{\rm in}=k)
\end{equation}

\noindent
is the generating function at $t=0$
and

\begin{equation}
t_s = \frac{t+s+1}{s+1}
\label{eq:t_s}
\end{equation}

\noindent
is a rescaled time, which is measured in units of the size of the seed
network (up to an adjustment of one step).
Using the series expansions

\begin{equation}
\frac{1}{x-2} = - \frac{1}{2} \sum_{k=0}^{\infty}  \left( \frac{x}{2} \right)^k,
\end{equation}

\noindent
and

\begin{equation}
t_s^{x-2} = 
\frac{1}{t_s^2}
\sum_{k=0}^{\infty}  \frac{( \ln t_s
)^k}{k!} x^k,
\end{equation}

\noindent
we express the generating function $G_t^{\rm in}(x)$ in the form

\begin{eqnarray}
G_t^{\rm in}(x) 
&=&
\frac{1}{t_s^2}
\sum_{k=0}^{\infty} x^k \sum_{m=0}^k \left[ P_0(K_{\rm in}=m) - \frac{1}{2^{m+1}} \right]
\frac{ ( \ln t_s )^{k-m} }{ (k-m)! }
\nonumber \\
&+&
\sum_{k=0}^{\infty} \frac{x^k}{2^{k+1}}.
\end{eqnarray}

\noindent
Therefore, the in-degree distribution of the backbone tree at time $t$ is given by

\begin{equation}
P_t(K_{\rm in}=k) = 
\frac{1}{t_s^2}
\sum_{m=0}^k \left[ P_0(K_{\rm in}=m) - \frac{1}{2^{m+1}} \right]
\frac{ ( \ln t_s )^{k-m} }{ (k-m)! }
+ \frac{1}{2^{k+1}}.
\label{eq:Ptkinbt}
\end{equation}

\noindent
At $t=0$ the only non-zero term of the sum in Eq. (\ref{eq:Ptkinbt})
is $m=k$, and the expression for the in-degree distribution
$P_t(K_{\rm in}=k)$ is reduced to $P_0(K_{\rm in}=k)$.
At short times $P_t(K_{\rm in}=k)$ is still dominated by its initial
condition, $P_0(K_{\rm in}=k)$, whose weight is diminished as $t$
increases.

In the long time limit, 
the first term on the right hand side of Eq. (\ref{eq:Ptkinbt}) vanishes
and the in-degree distribution of the backbone tree converges towards an 
asymptotic form, which is given by

\begin{equation}
P(K_{\rm in}=k) = 
\frac{1}{2^{k+1}}.
\label{eq:PbKlt}
\end{equation}

\noindent
This implies that in the long time limit
$P(K_{\rm in}=0)=1/2$,
namely half of the nodes in the network are
source nodes that have no incoming links.
Interestingly, this result carries over to the 
corded DND network for any value of $0 \le p \le 1$.

The out-degree distribution of the backbone tree,
denoted by
$P_t(K_{\rm out}=k)$,
evolves in time according to

\begin{equation}
P_{t+1}(K_{\rm out}=k) = 
\frac{ 
 N_t P_t(K_{\rm out}=k) 
+ \delta_{k,1} }{N_t+1}.
\label{eq:recK0}
\end{equation}

\noindent
The first term on the right hand side accounts for the out-degrees of the
existing nodes in the network, which are not affected by the addition of
the new node D.
The second term accouts for the out-degree of the daughter node D
(which is $k=1$).
Subtracting $P_t(K_{\rm out}=k)$ from both sides 
of Eq. (\ref{eq:recK0})
and replacing the difference on the left
hand side by a time derivative we obtain

\begin{equation}
\frac{\partial}{\partial t} P_t(K_{\rm out}=k) = 
- \frac{ P_t(K_{\rm out}=k) 
- \delta_{k,1} }{t+s+1}.
\label{eq:mera}
\end{equation}

\noindent
Solving Eq. (\ref{eq:mera}) we obtain the
out-degree distribution of the 
backbone tree, which  
is given by

\begin{equation}
P_t(K_{\rm out}=k) = \frac{  (s+1) P_0(K_{\rm out}=k) + t \delta_{k,1}  }{t+s+1}.
\label{eq:PbK}
\end{equation}

\noindent
In the long time limit of $t \gg s$ the out-degree distribution converges
towards

\begin{equation}
P_t(K_{\rm out}=k) = \delta_{k,1},
\end{equation}

\noindent
as expected for a tree structure.

Since nodes are discrete entities the
node duplication is an intrinsically discrete process.
Therefore, the replacement of the difference
$P_{t+1}(K_{\rm out}=k) - P_t(K_{\rm out}=k)$
on the left hand side of Eq. (\ref{eq:mera})
by $\partial P_t(K_{\rm out}=k)/ \partial t$
is an approximation.
In fact, it is closely related to the approximation involved in numerical
integration of differential equations using the Euler method.
In the Euler method the time derivative $df/dt$ is replaced by
$(f_{t+\Delta t}-f_t)/\Delta t$, where $\Delta t$ is a suitable chosen
time step. In our case $\Delta t=1$.
Below we evaluate the error associated with this approximation. 
This error can be expressed by a series expansion of the form

\begin{eqnarray}
P_{t+1}(K_{\rm out}=k) - P_t(K_{\rm out}=k) &=&
\frac{\partial}{\partial t} P_t(K_{\rm out}=k)
\nonumber \\
&+&
\frac{1}{2}
\frac{\partial^2}{\partial t^2} P_t(K_{\rm out}=k)
+ \dots,
\end{eqnarray}

\noindent
where

\begin{equation}
\frac{1}{2}
\frac{\partial^2}{\partial t^2} P_t(K_{\rm out}=k)
=
\frac{ P_t(K_{\rm out}=k) 
- \delta_{k,1} }{(t+s+1)^2}.
\end{equation}

\noindent
Thus, the leading correction term to Eq. (\ref{eq:mera}) scales
like $1/(t+s+1)^2$ and quickly vanishes as the growth process evolves
and the network size increases.
It can be shown that a similar scaling of the correction term is obtained
for the in-degree and out-degree distributions of the whole network
as well as for the distributions of the numbers of upstream
and downstream nodes.
This means that the replacement of the difference by a time derivative
has little effect on the results.

\section{The number of upstream nodes}

Consider a random node $i$, in the corded DND network. 
As in other directed networks, $i$ can be reached along directed paths
from only some of the nodes in the network.
These nodes are referred to as upstream nodes of $i$.
We denote the number of upstream nodes of $i$ by $N_{\rm up}^{i}$.
Below we calculate the distribution $P_t(N_{\rm up}=n)$ of the
number of upstream nodes of random nodes in the network.
At each time step a new node D is added to the network.
The in-degree of the newly added node is $k^{\rm D}_{\rm in}=0$. Therefore,
the number of upstream nodes of D, 
upon its formation
at time $t$, is $N_{\rm up}^{\rm D}=0$. 
This property can be expressed by 
$P_t(N_{\rm up}^{\rm D}=n) = \delta_{n,0}$.
Consider a node, $i$, which is an outgoing neighbor of the mother node, M.
Clearly, M is an upstream node of $i$. 
Since there is a directed link from D to M, the daughter node D
is also an upstream node of $i$. Thus, due to the duplication of M 
each one of 
its outgoing neighbor nodes $i$ gains one upstream node, namely
$N_{\rm up}^{i} \rightarrow N_{\rm up}^{i}+1$.
Therefore, the distribution of the number of upstream nodes evolves according to

\begin{eqnarray}
P_{t+1}(N_{\rm up} = n)  &=&
 \frac{ n P_t(N_{\rm up} = n-1) }{N_t+1}
\nonumber \\
&+& \frac{ (N_t-n-1) P_t(N_{\rm up} = n) + \delta_{n,0} }{N_t+1}.
\end{eqnarray}

\noindent
Subtracting $P_t(N_{\rm up}=n)$
from both sides
and replacing the difference on the left hand side by a time derivative,
we obtain

\begin{equation}
\frac{\partial}{\partial t} P_t(N_{\rm up} =n) =
 \frac{ n P_t(N_{\rm up} = n-1) - (n+2) P_t(N_{\rm up} = n) + \delta_{n,0} }{t+s+1}.
\label{eq:dPNupt}
\end{equation}

\noindent
Writing Eq. (\ref{eq:dPNupt}) in terms of the generating function

\begin{equation}
G_t^{\rm up}(x) = \sum_{n=0}^{\infty}
x^n P_t(N_{\rm up}=n),
\end{equation}

\noindent
we obtain

\begin{equation}
\frac{\partial}{\partial t} G_t^{\rm up}(x) =
\frac{1}{t+s+1} 
\left[ x(x-1) \frac{\partial}{\partial x} G_t^{\rm up}(x) + (x-2) G_t^{\rm up}(x) +1 \right].
\end{equation}

\noindent
The solution of this equation is

\begin{eqnarray}
G_t^{\rm up}(x) &=& 
\frac{(s+1)^2}{(t+s+1) [s + 1 + t(1-x)]}
G_0^{\rm up} \left[ \frac{(s+1) x}{s+1 + t(1-x)} \right]
\nonumber \\
&+& 
\frac{t}{(t+s+1)x}  
- \frac{1-x}{x^2} \ln \left[ \frac{t+s+1}{s+1 + t(1-x)} \right].
\label{eq:Gtupx}
\end{eqnarray}

\noindent
Note that in spite of the apparent singularity 
of the last two terms on the right hand side of Eq. (\ref{eq:Gtupx})
at $x=0$, a careful expansion near the origin reveals that 
the sum of these two terms is perfectly regular there.
Also, one can easily confirm that at $t=0$ the generating function
$G_t^{\rm up}(x)$ is reduced to $G_0^{\rm up}(x)$.

The distribution $P_t(N_{\rm up}=n)$ can be extracted from $G_t^{\rm up}(x)$
using the relation

\begin{equation}
P_t(N_{\rm up}=n) = \frac{1}{n!} \frac{\partial^n}{\partial x^n} G_t^{\rm up}(x) \bigg\vert_{x=0}.
\end{equation}

\noindent
Carrying out the derivatives, we obtain

\begin{eqnarray}
P_t(N_{\rm up}=n) &=&
\left( \frac{t}{t+s+1} \right)^{n+2}
\sum_{m=0}^{n}
{ {n} \choose {m} } \left( \frac{s+1}{t} \right)^{m+2} P_0(N_{\rm up}=m)
\nonumber \\
&+&
\frac{1}{n+1} \left( \frac{t}{t+s+1} \right)^{n+1}
-
\frac{1}{n+2} \left( \frac{t}{t+s+1} \right)^{n+2}.
\label{eq:PNupt}
\end{eqnarray}

\noindent
In the long time limit, $t \gg s$, the generating function
converges to the asymptotic form

\begin{equation}
G^{\rm up}(x) = \frac{1}{x}
+ \frac{1-x}{x^2} \ln (1-x).
\end{equation}

\noindent

The asymptotic form of the generating function can be
expressed by

\begin{equation}
G^{\rm up}(x) = 
\sum_{n=0}^{\infty}
\frac{x^n}{(n+1)(n+2)}.
\end{equation}

\noindent
Therefore, in the asymptotic limit

\begin{equation}
P(N_{\rm up}=n) = \frac{1}{(n+1)(n+2)},
\label{eq:PNup}
\end{equation}

\noindent
where $n=0,1,\dots,N_t-1$.
Thus, in the long time limit the distribution of the number of upstream
nodes converges towards a power-law distribution whose tail is
given by

\begin{equation}
P(N_{\rm up}=n) \sim \frac{1}{n^{\alpha}},
\label{eq:PnupPL}
\end{equation}

\noindent
where $\alpha=2$.

The mean number of upstream nodes is given by

\begin{equation}
\langle N_{\rm up} \rangle_t= \sum_{n=0}^{t+s-1}
n P_t(N_{\rm up}=n),
\end{equation}

\noindent
where $P_t(N_{\rm up}=n)$ is given by Eq. (\ref{eq:PNupt}).
Carrying out the summation, it is found that

\begin{equation}
\langle N_{\rm up} \rangle_t = \langle N_{\rm up} \rangle_0 +  \ln t_s.
\label{eq:Nupm}
\end{equation}

\noindent
Thus, in the long time limit the mean number of upstream nodes
scales with time like

\begin{equation}
\langle N_{\rm up} \rangle_t =   \ln t .
\label{eq:Nupmlong}
\end{equation}

The second factorial moment is given by

\begin{equation}
\langle N_{\rm up} (N_{\rm up}-1) \rangle_t = \frac{ \partial^2}{\partial x^2} 
G_t^{\rm up}(x) \bigg\vert_{x=1}.
\end{equation}

\noindent
Carrying out the derivatives, we obtain

\begin{eqnarray}
\langle N_{\rm up} (N_{\rm up}-1) \rangle_t 
&=&
\langle N_{\rm up}(N_{\rm up}-1) \rangle_0  t_s  
\nonumber \\
&+& 4 \frac{t}{s+1} ( \langle N_{\rm up} \rangle_0 +1 )
- 4 \ln t_s.
\end{eqnarray}

\noindent
Therefore, the second moment of $P_t(N_{\rm up}=n)$ is 

\begin{equation}
\langle N_{\rm up}^2 \rangle_t = 
\langle N_{\rm up}^2 \rangle_0  
+ 
( 4 + 3 \langle N_{\rm up} \rangle_0  + \langle N_{\rm up}^2 \rangle_0   )\frac{t}{s+1}
-
3 \ln t_s.
\label{eq:Nup}
\end{equation}

\noindent
Thus, in the long time limit the second moment
$\langle N_{\rm up}^2 \rangle_t$
increases linearly with the network size.
The variance is given by

\begin{eqnarray}
{\rm Var}_t(N_{\rm up}) &=& 
{\rm Var}_0(N_{\rm up})
+ ( 4 + 3 \langle N_{\rm up} \rangle_0 + \langle N_{\rm up}^2 \rangle_0 )  \frac{t}{s+1}
\nonumber \\
&-&
\left( 2 \langle N_{\rm up} \rangle_0 + 3 \right) \ln t_s
- ( \ln t_s )^2.
\label{eq:Varnup}
\end{eqnarray}

\noindent
In the long time limit the variance diverges like

\begin{equation}
{\rm Var}_t(N_{\rm up}) \simeq
\left( \frac{4 + 3 \langle N_{\rm up} \rangle_0 + \langle N_{\rm up}^2 \rangle_0}{s+1} \right)  t.
\label{eq:Varnuplt}
\end{equation}

\noindent
In the case in which the seed network consists of a single node,
the variance is reduced to

\begin{equation}
{\rm Var}_t(N_{\rm up}) \simeq 2 t.
\end{equation}

\noindent
Thus, in the long time limit the mean number of upstream nodes
diverges logarithmically with the network size while the variance
of $P(N_{\rm up}=n)$ diverges linearly with the network size.
This reflects the fact that the power-law distribution of Eq. (\ref{eq:PnupPL})
with $\alpha=2$ is a marginal case between $\alpha > 2$ where the first
moment is finite and $\alpha \le 2$ where the first moment diverges.

\section{The number of downstream nodes}

Consider a random node $i$ in the corded DND network.
As in other directed networks, only some of the nodes in
the network can be reached from $i$ along directed paths. 
In the corded DND network, 
all the nodes that can be reached from $i$ are older than $i$
and reside on directed paths from $i$ to the sink node.
Below we calculate the distribution 
$P_t(N_{\rm down}=n)$,
of the number of nodes that can
be reached along directed paths from a random node $i$ at time $t$.
To this end we derive a recursion equation for
$P_t(N_{\rm down}=n)$ 
during the network growth process 
starting from a seed network of $s$ nodes
in which the distribution of the number of downstream nodes
is given by $P_0(N_{\rm down}=n)$,
where $n=0,1,\dots,s-1$.
Clearly, $P_0(N_{\rm down}=n)=0$ for $n \ge s$.

At time $t$ we pick a random mother node, M.
The number of nodes that can be reached from node 
$M$ is denoted by $N_{\rm down}^{\rm M}$.
Since M is chosen randomly,
$P_t(N^{\rm M}_{\rm down}=n) = P_t(N_{\rm down}=n)$.
The number of nodes that can be reached from the daughter
node, D, is thus $N^{\rm D}_{\rm down} = N^{\rm M}_{\rm down} + 1$, because it
includes the node M itself and all its downstream nodes.
Thus, the distribution 
$P_t(N^{\rm D}_{\rm down}=n)$ 
of the number of nodes that can be
reached from node D, upon its formation at time $t$,
is given by

\begin{equation}
P_t(N^{\rm D}_{\rm down}=n) = P_t(N_{\rm down}=n-1).
\label{eq:NoD}
\end{equation}

\noindent
Therefore,

\begin{equation}
P_{t+1}(N_{\rm down}=n) = 
\frac{ N_t P_t(N_{\rm down}=n) 
+ 
P_t(N^{\rm D}_{\rm down}=n) }{N_t+1}.
\label{eq:Nt1}
\end{equation}

\noindent
Inserting Eq. (\ref{eq:NoD}) into Eq. (\ref{eq:Nt1}),
subtracting $P_t(N_{\rm down}=n)$ 
from  both sides
and replacing the difference on the left hand side by a time
derivative, we obtain

\begin{equation}
\frac{\partial}{\partial t} P_t(N_{\rm down}=n) =
- \frac{ P_t(N_{\rm down}=n) - P_t(N_{\rm down}=n-1) }{t+s+1}.
\label{eq:Nt2}
\end{equation}

\noindent
Expressing Eq. (\ref{eq:Nt2}) in terms of the generating function

\begin{equation}
G_t^{\rm down}(x) = \sum_{n=0}^{\infty}
x^n P_t(N_{\rm down}=n),
\end{equation}

\noindent
we obtain

\begin{equation}
\frac{\partial}{\partial t} G_t^{\rm down}(x) =
\frac{x-1}{t+s+1} G_t^{\rm down}(x).
\end{equation}

\noindent
The solution of this equation is

\begin{equation}
G_t^{\rm down}(x) = 
\frac{1}{t_s^{1-x}} G_0^{\rm down}(x),
\label{eq:Gtdown}
\end{equation}

\noindent
where
$G_0^{\rm down}(x)$
is the generating function of the
distribution of the number of downstream nodes in
the seed network, $P_0(N_{\rm down}=n)$.
Rewriting Eq. (\ref{eq:Gtdown}) as a series expansion in
powers of $x$, we find that

\begin{equation}
P_t(N_{\rm down}=n) = \frac{1}{t_s}
\sum_{m=0}^{n} \frac{ P_0(N_{\rm down}=m) (\ln t_s)^{n-m} }{(n-m)!}.
\end{equation}

\noindent
In the case that the seed network consists of a single node,
$s=1$ and $P_0(N_{\rm down}=m)=\delta_{m,0}$.
In this case the distribution of the number of downstream nodes
is reduced to a Poisson distribution of the form

\begin{equation}
P_t(N_{\rm down}=n) = 
\frac{ e^{- \ln (t_s)} (\ln t_s)^{n} }{n!},
\end{equation}

\noindent
where $t_s = (t+2)/2$
and 
$\langle N_{\rm down} \rangle_t = \ln(t_s)$.

The mean number of downstream nodes is given by

\begin{equation}
\langle N_{\rm down} \rangle_t = \frac{ \partial}{\partial x} 
G_t^{\rm down}(x) \bigg\vert_{x=1}.
\end{equation}

\noindent
Carrying out the differentiation, we obtain

\begin{equation}
\langle N_{\rm down} \rangle_t = 
\langle N_{\rm down} \rangle_0
+ \ln t_s.
\label{eq:Ndown}
\end{equation}

\noindent
Thus, in the long time limit the mean number of downstream nodes scales like

\begin{equation}
\langle N_{\rm down} \rangle_t = 
\ln t.
\label{eq:Ndownlongt}
\end{equation}

\noindent
Note that the mean number of downstream nodes, given
by Eq. (\ref{eq:Ndown}), is equal to the mean number of
upstream nodes, given by Eq. (\ref{eq:Nupm}), for any time, $t$
(including the seed network, for which $t=0$).
This is due to the fact that each directed link, from node $i$ 
to node $j$, contributes one upstream node to $j$ and one
downstream node to $i$.

The second factorial moment is given by

\begin{equation}
\langle N_{\rm down} (N_{\rm down}-1) \rangle_t = \frac{ \partial^2}{\partial x^2} 
G_t^{\rm down}(x) \bigg\vert_{x=1}.
\end{equation}

\noindent
Carrying out the differentiation, we obtain

\begin{eqnarray}
\langle N_{\rm down} (N_{\rm down}-1) \rangle_t 
&=&
\langle N_{\rm down} (N_{\rm down}-1) \rangle_0
\nonumber \\
&+& 2 \langle N_{\rm down} \rangle_0 \ln t_s
+ (\ln  t_s)^2.
\end{eqnarray}

\noindent
Therefore, the second moment of $P_t(N_{\rm down}=n)$ is given by

\begin{equation}
\langle N_{\rm down}^2 \rangle_t =
\langle N_{\rm down}^2  \rangle_0
+ \left( 2 \langle N_{\rm down} \rangle_0 + 1 \right) \ln t_s
+ \ln^2 t_s.
\end{equation}

\noindent
The variance of $P_t(N_{\rm down}=n)$ is given by

\begin{equation}
{\rm Var}_t(N_{\rm down}) =
{\rm Var}_0(N_{\rm down})
+ \ln t_s,
\label{eq:VarNdown}
\end{equation}

\noindent
which in the long time limit converges towards

\begin{equation}
{\rm Var}_t(N_{\rm down}) =
\ln t.
\label{eq:VarNdownlongt}
\end{equation}

\noindent
We thus find that in the long time limit 
$\langle N_{\rm down} \rangle_t = {\rm Var}_t(N_{\rm down}) = \ln t$,
which is consistent with the fact that in this limit the distribution of the number of downstream
nodes converges towards a Poisson distribution.

\section{The mean degree}

While in the sections above we focused on properties of the backbone tree,
here we start to analyze the structure of the whole DND network.
Consider an ensemble of corded DND networks of size $N_t=t+s$.
The mean degree of such a network ensemble 
is denoted by 
$\langle K \rangle_t  = \langle K_{\rm in} \rangle_t = \langle K_{\rm out} \rangle_t $.
Selecting a random node, M, for duplication at time
$t$, the expected out-degree of the daughter node is

\begin{equation}
\langle K^{\rm D}_{\rm  out} \rangle_t = 1 + p \langle K_{\rm out} \rangle_t.
\end{equation}

\noindent
After the duplication step is completed, the
mean out-degree of the network at time $t+1$ 
is given by

\begin{equation}
\langle K_{\rm out} \rangle_{t+1} =
\frac{N_t \langle K_{\rm out} \rangle_t +  \langle K^{\rm D}_{\rm out} \rangle_t}{N_t+1}.
\end{equation}

\noindent
Expressing 
$\langle K^{\rm D}_{\rm out} \rangle_t$
in terms of
$\langle K_{\rm out} \rangle_t$
we obtain

\begin{equation}
\langle K_{\rm out} \rangle_{t+1} =
\frac{(N_t+p) \langle K_{\rm out} \rangle_t + 1}{N_t+1}.
\end{equation}

\noindent
Subtracting $\langle K_{\rm out} \rangle_t$
from both sides and replacing the difference on
the left-hand side by a time derivative we obtain

\begin{equation}
\frac{d}{dt} \langle K_{\rm out} \rangle_t =
\frac{(p-1) \langle K_{\rm out} \rangle_t + 1}{t+s+1}.
\label{eq:dkdt}
\end{equation}

\noindent
Starting from a seed network of $s$ nodes whose mean out-degree  
is given by $\langle K_{\rm out} \rangle_0$,
the solution of Eq. 
(\ref{eq:dkdt})
takes the form

\begin{equation}
\langle K_{\rm out} \rangle_t =  
\left( \langle K_{\rm out} \rangle_0 - \frac{1}{1-p} \right) 
\frac{1}{ t_s^{1-p} }
+
\frac{1}{1-p}.
\label{eq:k}
\end{equation}

\noindent
At $t=0$ the rescaled time is $t_s=1$, while in the long time limit
$t_s \simeq t/(s+1)$.
In the long limit the 
first term on the right hand side of Eq. (\ref{eq:k}) vanishes
and the 
mean out-degree converges to

\begin{equation}
\langle K_{\rm out} \rangle
= \frac{1}{1-p}.
\label{eq:kinf}
\end{equation}

\noindent
As time proceeds, $\langle K_{\rm out} \rangle_t$ converges to its asymptotic values,
$\langle K_{\rm out} \rangle$, such that 
$\langle K_{\rm out} \rangle_t - \langle K_{\rm out} \rangle \sim 1/t_s^{1-p}$.
This power-law dependence on $t_s$ indicates that the convergence is slow,
particularly in the limit of $p \rightarrow 1$. The convergence rate can be
characterized by the time it takes the difference
$\langle K_{\rm out} \rangle_t - \langle K_{\rm out} \rangle$ 
to be reduced to $1/e$ of its
initial value. 
This time, which is given by 

\begin{equation}
\tau = (s+1) \left[ \exp \left( \frac{1}{1-p} \right)  - 1 \right],
\end{equation}

\noindent
can be used as a crossover time between the short-time regime,
in which $t \ll \tau$ and the long-time regime, in which $t \gg \tau$.
The crossover time $\tau$ is proportional to the size of the seed network.
Since $\langle K_{\rm in} \rangle_t = \langle K_{\rm out} \rangle_t$,
the mean of the in-degree distribution at time $t$ is also given by Eq.
(\ref{eq:k}) and in the asymptotic limit by Eq. (\ref{eq:kinf}).

\section{The in-degree distribution}

The in-degree distribution
$P_t(K_{\rm in}=k)$,
of the corded DND network, evolves as the network grows.
As a daugther node, D, is added to the network, the in-degrees of the
mother node and some of its outgoing neighbors increase by $1$.
The in-degree of the daughter node itself is $k^{\rm D}_{\rm in}=0$,
namely
$P_t(K^{\rm D}_{\rm in}=k)=\delta_{k,0}$.
The in-degree distribution of the existing nodes, 
at time $t+1$ is given by

\begin{eqnarray}
P_{t+1}(K_{\rm in} = k) 
&=& 
\frac{ 1+(k-1)p }{N_t} P_t(K_{\rm in}=k-1)
\nonumber \\
&+& \frac{  N_t - (1+kp) }{N_t}  P_t(K_{\rm in}=k).
\end{eqnarray}

\noindent
Taking into account the contribution of the newly formed node D,
with a suitable normalization, we obtain

\begin{eqnarray}
P_{t+1}(K_{\rm in}=k) 
&=&
\frac{1+(k-1)p }{N_t + 1} P_t(K_{\rm in}=k-1) 
\nonumber \\
&+&
\frac{ N_t - (1+kp) }{N_t + 1} P_t(K_{\rm in}=k) 
+
\frac{ \delta_{k,0} }{N_t+1}.
\end{eqnarray}

\noindent
Here the denominator is $N_t+1$ rather than $N_t$ in order to 
account for the new node.
Subtracting $P_t(K_{\rm in}=k)$ from both sides 
and replacing the difference on the left-hand side by a time derivative,
we obtain

\begin{eqnarray}
\frac{\partial}{\partial t} P_t(K_{\rm in}=k) 
&=&
\frac{ 1+(k-1)p }{t+s+1} P_t(K_{\rm in}=k-1) 
\nonumber \\
&-& 
\frac{ 2+kp }{t+s+1} P_t(K_{\rm in}=k) 
+
\frac{ \delta_{k,0} }{t+s+1}.
\end{eqnarray}

\noindent
Expressing this equation in terms of the
generating function

\begin{equation}
G_t^{\rm in}(x) = \sum_{k=0}^{\infty} x^k P_t(K_{\rm in}=k)
\end{equation}

\noindent
we obtain

\begin{equation}
\frac{\partial}{\partial t} G_t^{\rm in}(x) =
\frac{1}{t+s+1}
\left[ px(x-1) \frac{\partial}{\partial x} G_t(x)
+
(x-2) G_t(x) +1 \right],
\label{eq:Gtin1}
\end{equation}

\noindent
with the initial condition

\begin{equation}
G_0^{\rm in}(x) = \sum_{k=0}^{\infty} x^k P_0(K_{\rm in}=k).
\end{equation}

\noindent
The solution of Eq. (\ref{eq:Gtin1}) is 

\begin{eqnarray}
G_t^{\rm in}(x) &=& 
\frac{1}{t_s}
\left[ \frac{1}{x + (1-x)   
t_s^p 
} \right]^{\frac{1}{p}}
G_0^{\rm in} \left[ \frac{x}{x + (1-x) 
t_s^p
} \right]
\nonumber \\
&-& \frac{1}{2} \left[ \frac{\sqrt{1-x}}{x +(1-x)
t_s^p} 
\right]^{ \frac{2}{p} }  
{_2F_1}\left[ 1+\frac{1}{p}, \frac{2}{p} ; 1 + \frac{2}{p} ; \frac{ x}{ x+(1-x)
t_s^p
} 
\right]
\nonumber \\
&+&
\frac{1}{2} \   {_2F_1} \left( 1, \frac{1}{p} ; 1 + \frac{2}{p} ; x \right),
\label{eq:Gtin}
\end{eqnarray}

\noindent
where 
$t_s$ is given by Eq. (\ref{eq:t_s})
and
$_2F_1(a,b,c,z)$ is the hypergeometric function
\cite{Olver2010}.
In the long time limit, 
$t \rightarrow \infty$, the first two terms on the right hand side of Eq. (\ref{eq:Gtin})
vanish and 
$G_t^{\rm in}(x)$ 
converges towards an asymptotic form which is given by

\begin{equation}
G^{\rm in}(x) = \frac{1}{2} \ {_2}F_1 \left( 1, \frac{1}{p}; 1+\frac{2}{p}; x \right).
\end{equation}

\noindent
Expressing $G^{\rm in}(x)$ using a series expansion in powers of $x$, we obtain

\begin{equation}
G^{\rm in}(x) = \sum_{k=0}^{\infty} 
\frac{ \Gamma \left(\frac{2}{p} \right) \Gamma \left( \frac{1}{p} + k \right) }
{ (2+pk) \Gamma \left( \frac{1}{p} \right) 
\Gamma \left( \frac{2}{p} + k \right) }
x^k,
\label{eq:Gtxin}
\end{equation}

\noindent
where $\Gamma(z)$ is the Gamma function.
Eq. (\ref{eq:Gtxin}) implies that
the in-degree distribution in the long time limit is given by

\begin{equation}
P(K_{\rm in}=k) = \frac{ \Gamma \left(\frac{2}{p} \right) \Gamma \left( \frac{1}{p} + k \right) }
{ (2+pk) \Gamma \left( \frac{1}{p} \right) \Gamma \left( \frac{2}{p} + k \right) }.
\label{eq:PtKins}
\end{equation}

\noindent
In the asymptotic limit of large $k$, one can use the expansion

\begin{equation}
\frac{ \Gamma \left( \frac{1}{p} + k \right) }{ \Gamma \left( \frac{2}{p} + k \right) }
\simeq \frac{ p^{1/p} }{(2+pk)^{1/p} }
\left[ 1 + \frac{1+p}{2 p^2 k} + \mathcal{O} \left( \frac{1}{k^2} \right) \right].
\end{equation}

\noindent
Therefore, in the long time limit the tail of the in-degree distribution
is given by

\begin{equation}
P(K_{\rm in}=k) =
\frac{ \Gamma \left( \frac{2}{p} \right) }{ \Gamma \left( \frac{1}{p} \right) }
\frac{ p^{1/p} }{(2+pk)^{ 1 + 1/p} }.
\label{eq:PtKin}
\end{equation}

\noindent
The tail of the in-degree distribution thus follows a power-law distribution
of the form 

\begin{equation}
P(K_{\rm in}=k) \sim k^{- \gamma},
\end{equation}

\noindent
where 

\begin{equation}
\gamma = 1 + \frac{1}{p}.
\label{eq:gammap}
\end{equation}

\noindent
From Eq. (\ref{eq:PtKin}) it is found that the crossover from the behavior at low degrees
to the power-law tail takes place at $k \simeq 2/p$.
Thus, the lower cutoff of the power-law behavior is at

\begin{equation}
k_{\rm lower\_cutoff} \simeq \frac{2}{p}.
\end{equation}

\noindent
For a finite network of $N_t$ nodes the distribution $P(K_{\rm in}=k)$,
given by Eq. (\ref{eq:PtKin}) is bounded by an upper cutoff,
which is given by $P(K_{\rm in}=k) \simeq 1/N_t$.
Using the Stirling approximation to evaluate $\Gamma(2/p)$ and $\Gamma(1/p)$
we find that the upper cutoff is

\begin{equation}
k_{\rm upper\_cutoff} \simeq \frac{N_t^{\frac{p}{p+1}}}{p}.
\end{equation}

\noindent
The exponent $\gamma$, 
given by Eq. (\ref{eq:gammap}),
satisfies the condition 
$\gamma > 2$ for the whole range of $0 < p < 1$, 
thus the first moment of $P(K_{\rm in}=k)$ 
is finite even in the infinite network limit.
For $p \ge 1/2$, the exponent $\gamma$ satisfies $2 <\gamma \le 3$,
which means that 
in the infinite network limit
the second moment 
of $P(K_{\rm in}=k)$
diverges,
while for $p < 1/2$, $\gamma > 3$ and the second moment is finite.

In the limit of $p \rightarrow 0^{+}$, 
it can be shown that
the in-degree distribution presented by 
Eq. (\ref{eq:PtKins}) 
is reduced to the 
in-degree distribution of the backbone tree, 
given by Eq. (\ref{eq:PbKlt}).
This is done by using the asymptotic expansion of 
$\Gamma(x)$ in the limit $x = 1/p \rightarrow \infty$,
where $\ln \Gamma(x) \simeq x \ln x - x$.
Using this expansion we find that in the limit of
$p \rightarrow 0^{+}$

\begin{equation}
\frac{ \Gamma \left(\frac{2}{p} \right) \Gamma \left( \frac{1}{p} + k \right) }
{ \Gamma \left( \frac{1}{p} \right) \Gamma \left( \frac{2}{p} + k \right) }
\rightarrow \frac{1}{2^k}.
\end{equation}

\noindent
Inserting this result in Eq. 
(\ref{eq:PtKins}) for the asymptotic form of the in-degree distribution,
it coincides with 
Eq. (\ref{eq:PbKlt})
for the in-degree distribution of the backbone tree.
Similarly, in the limit of 
$p \rightarrow 0^{+}$
the generating function 
$G_t^{\rm in}(x)$ given by Eq. (\ref{eq:Gtin})
coincides with the corresponding generating function
of the backbone tree, given by Eq. (\ref{eq:Gtbbt}).
More specifically, as 
$p \rightarrow 0^{+}$
the first term in Eq. (\ref{eq:Gtin})
converges towards $G_0^{\rm in}(x)/t_s^{2-x}$,
the second term converges towards $- [(2-x) t_s^{2-x}]^{-1}$ and
the third term converges towards $1/(2-x)$.

In the limit of $p \rightarrow 1^{-}$,
at long times,
the in-degree distribution presented by
Eq. (\ref{eq:PtKins}) converges towards the
distribution of the number of upstream nodes. 
This is due to the fact that in this limit essentially all the upstream
nodes of a random node, $i$, form directed links to $i$.
In this limit, Eq. (\ref{eq:PtKins}) is reduced to the form
$P(K_{\rm in}=k) = 1/(k+2)^2$,
which in the large $k$ limit conincides with the distribution
$P(N_{\rm up}) = 1/[(n+1)(n+2)]$,
given by Eq. (\ref{eq:PNup}).

\begin{figure}
\centerline{
\includegraphics[width=6.5cm]{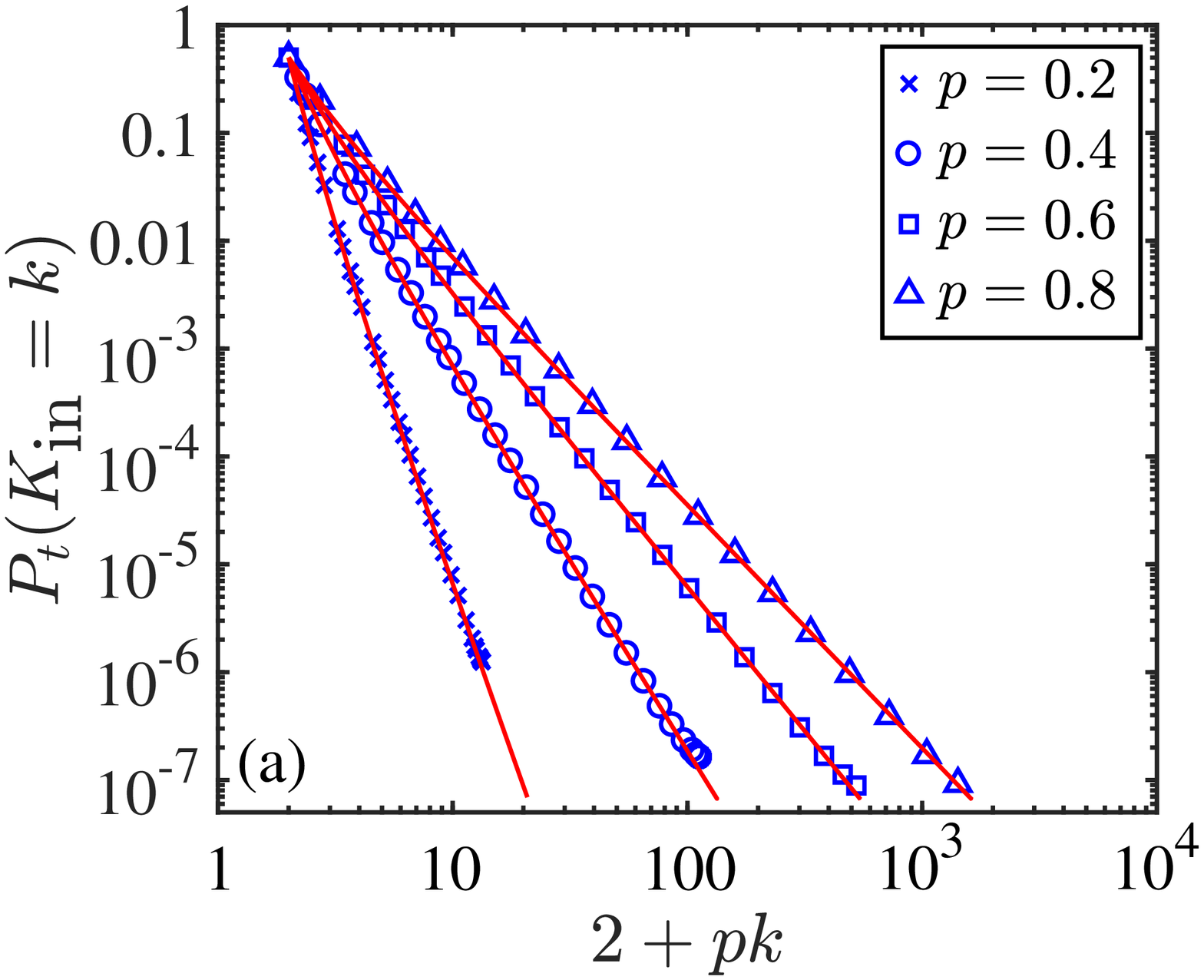}
\includegraphics[width=6.5cm]{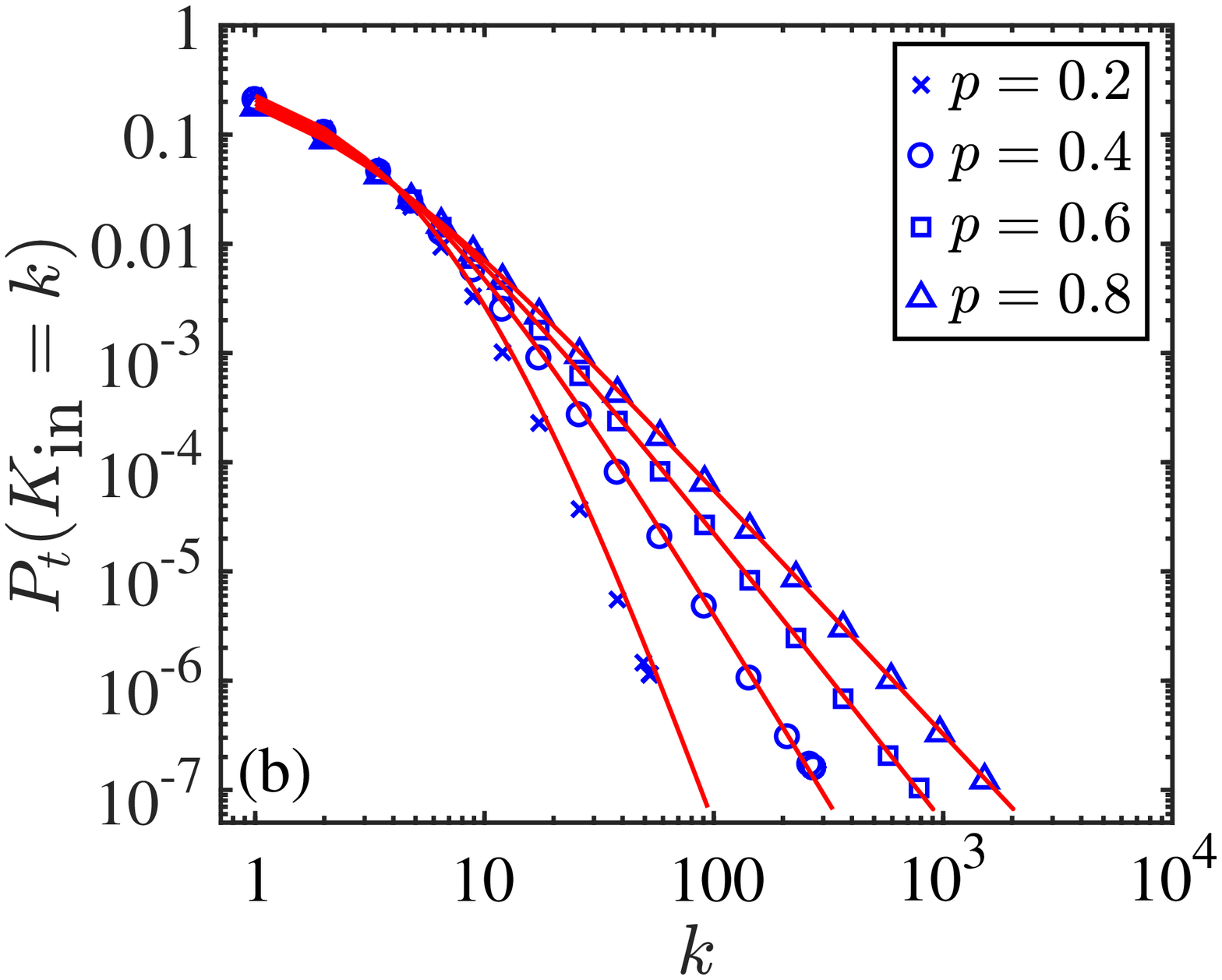}
}
\caption{
Analytical results (solid lines) for
the in-degree distribution
$P_t(K_{\rm in}=k)$, 
of the corded DND network
for $p=0.2$, $0.4$, $0.6$ and $0.8$,
on a log-log scale:
(a) as a function of $2+pk$;
and 
(b) as a function of $k$.
The analytical results are found to be in very good agreement with simulation
results (symbols).
The fact that the results shown in (a) are well fitted by straight lines
over the whole range confirms that the in-degree distribution follows
a shifted power-law distribution of the form
$P_t(K_{\rm in}=k) \sim (2+pk)^{-\gamma}$,
where the exponent is
given by $\gamma=1+1/p$.
The effect of the shift is apparent in (b) where there is a crossover from
a power-law tail for large $k$ to a non-scaling regime for small $k$.
}
\label{fig:3}
\end{figure}

In Fig. \ref{fig:3} we present analytical results (solid lines) for the in-degree distribution,
$P_t(K_{\rm in}=k)$, of the corded DND network, on a log-log scale,
as a function of $2+pk$ (a) and as a function of $k$ (b),
for $t=10^6$ and $p=0.2$, $0.4$, $0.6$ and $0.8$,
obtained from Eq. (\ref{eq:PtKins}).
The analytical results are found to be in excellent agreement with
the results obtained from computer simulations 
(symbols). The straight lines in Fig. \ref{fig:3}(a) confirm the result that
the in-degree distribution follows a shifted power-law distribution.

The mean of the in-degree distribution can be obtained by differentiating
the generating function, namely

\begin{equation}
\langle K_{\rm in} \rangle_t = \frac{\partial}{\partial x} G_t^{\rm in}(x) \bigg\vert_{x=1}.
\label{eq:Kinmean}
\end{equation}

\noindent
Carrying out the differentiation, we obtain

\begin{equation}
\langle K_{\rm in} \rangle_t 
= 
\left( \langle K_{\rm in} \rangle_0 - \frac{1}{1-p} \right)
\frac{1}{t_s^{1-p}}
+
\frac{1}{1-p},
\label{eq:Kinm}
\end{equation}

\noindent
where $t_s$ is given by Eq. (\ref{eq:t_s}),
matching the results of Eq. (\ref{eq:k}), 
in agreement with the condition that
$\langle K_{\rm in} \rangle = \langle K_{\rm out} \rangle$.
In the long time limit the first term on the right hand side of Eq. (\ref{eq:Kinm})
vanishes and $\langle K_{\rm in} \rangle$ converges to the asymptotic value
$\langle K_{\rm in} \rangle = 1/(1-p)$.
The second factorial moment of the degree distribution, in the long time limit, is
given by

\begin{equation}
\langle K_{\rm in}(K_{\rm in}-1) \rangle = 
\frac{\partial^2}{\partial x^2} G^{\rm in}(x) \bigg\vert_{x=1}.
\end{equation}

\noindent
Carrying out the differentiation, we obtain

\begin{equation}
\langle K_{\rm in} (K_{\rm in} - 1) \rangle 
=
\frac{ _2F_{1} \left(3,2+\frac{1}{p};3+\frac{2}{p};1 \right) }{2(2+p)}
=
\frac{2(1+p)}{(1-p)(1-2p)}.
\label{eq:KK1m}
\end{equation}

\noindent
Summing up Eqs. (\ref{eq:Kinm}) and (\ref{eq:KK1m}),
we obtain

\begin{equation}
\langle K_{\rm in}^2 \rangle = \frac{3}{(1-p)(1-2p)}.
\label{eq:Kin2}
\end{equation}

\noindent
The variance 
${\rm Var}(K_{\rm in}) = \langle K_{\rm in}^2 \rangle - \langle K_{\rm in} \rangle^2$,
in the long time limit, 
is thus given by

\begin{equation}
{\rm Var}(K_{\rm in}) = \frac{2-p}{(1-p)^2(1-2p)}.
\label{eq:Varkin}
\end{equation}

\noindent
Note that Eqs. (\ref{eq:KK1m})-(\ref{eq:Varkin}) for the
second moment and the variance of $P(K_{\rm in}=k)$ are valid
only for $0 \le p < 1/2$.
As $p$ approaches $1/2$ from below, 
the second moment 
$\langle K_{\rm in} \rangle$,
given by Eq. (\ref{eq:Kin2})
diverges.
This is due to the fact that
in the limit $p \rightarrow 1/2^{-}$,
the exponent $\gamma$, in Eq. (\ref{eq:PtKin}),
satisfies
$\gamma \rightarrow 3^{+}$.

Working out higher order derivatives, one can show that
the factorial moments are given by

\begin{equation}
\langle K_{\rm in} (K_{\rm in} - 1) \dots (K_{\rm in} -n+1) \rangle 
= 
\frac{\partial^n}{\partial x^n} G^{\rm in}(x) \bigg\vert_{x=1}.
\end{equation}

\noindent
Carrying out the differentiations we obtain

\begin{equation}
\langle K_{\rm in}(K_{\rm in}-1) \dots (K_{\rm in}-n+1) \rangle 
= 
\frac{n!}{1-np} \prod_{k=1}^{n-1} \frac{1+kp}{1-kp}.
\end{equation}

\noindent
The moments $\langle K_{\rm in}^n \rangle$ can be obtained from
the factorial moments, using the relation

\begin{equation}
\langle K_{\rm in}^n \rangle 
=
\sum_{r=0}^n {n \brace r}
\langle K_{\rm in} (K_{\rm in} - 1) \dots (K_{\rm in}-n+1) \rangle,
\end{equation}

\noindent
where

\begin{equation}
{ n \brace k }
=
\frac{1}{k!}
\sum_{r=0}^n (-1)^{k-r} 
{ {k} \choose {r} } r^n
\end{equation}

\noindent
is the Stirling number of the second kind,
which represents the number of ways to partition a set of $n$ objects into
$k$ non-empty subsets
\cite{Olver2010}.
Therefore, the $n^{\rm th}$ moment of the in-degree distribution is given by

\begin{equation}
\langle K_{\rm in}^n \rangle =
\sum_{r=0}^{n} \sum_{j=0}^{r}
(-1)^{r-j} 
{ {r} \choose {j} } j^n 
\frac{1}{1-rp}
\prod_{k=1}^{r-1}  \frac{1+kp}{1-kp}.
\end{equation}

\noindent
Note that the moment $\langle K_{\rm in}^n \rangle$ is bounded
only for values of $p$ in the range
$0 < p < 1/n$.

\section{The out-degree distribution}

While the in-degree of a node in the corded DND network may continue to
increase as new nodes are added to the network, the out-degree
is determined upon formation of the node.
The out-degree distribution, $P_t^{\rm D}(K_{\rm out}=k)$, 
of a newly formed daughter node at time $t$ is given by

\begin{equation}
P_t^{\rm D}(K_{\rm out}=k) 
=
\sum_{m=k-1}^{\infty}  
{m \choose {k-1}} p^{k-1} (1-p)^{m-k+1}
P_t(K_{\rm out}=m).
\end{equation}

\noindent
Taking into account the contribution of D to the out-degree distribution of the
network we obtain

\begin{eqnarray}
P_{t+1}(K_{\rm out}=k) &=& 
\frac{  N_t P_t(K_{\rm out}=k) }{N_t+1} 
\nonumber \\
&+&
\frac{ 
\sum_{m=k-1}^{\infty}  {m \choose {k-1}} p^{k-1} (1-p)^{m-k+1}
P_t(K_{\rm out}=m) }{N_t + 1}.
\end{eqnarray}

\noindent
Subtracting $P_t(K_{\rm out}=k)$ from both sides and replacing the
difference on the left hand side by a time derivative, we obtain

\begin{eqnarray}
\frac{\partial}{\partial t} P_t(K_{\rm out}=k) 
&=&
\frac{  - P_t(K_{\rm out}=k) }{t+s+1}
\nonumber \\
&+&
\frac{
\sum_{m=k-1}^{\infty}  {m \choose {k-1}} p^{k-1} (1-p)^{m-k+1}
P_t(K_{\rm out}=m)  
}{t+s+1}.
\label{eq:dPdtKout}
\end{eqnarray}

\noindent
Assuming that $P_t(K_{\rm out}=k)$ converges to a steady state
such that in the long time limit 
$dP_t(K_{\rm out}=k)/dt=0$,
one obtains an equation
for the asymptotic distribution $P(K_{\rm out}=k)$,
which takes the form

\begin{equation}
P(K_{\rm out}=k)
= 
\sum_{m={k-1}}^{\infty}  
{ {m} \choose {k-1} } 
p^{k-1} (1-p)^{m-k+1} P(K_{\rm out}=m).
\label{eq:steady}
\end{equation}

\noindent
The generating function of the out-degree distribution,
in the long time limit,
is given by

\begin{equation}
G^{\rm out}(x) = \sum_{k=0}^{\infty}
x^k P(K^{\rm out}=k) .
\label{eq:Gxout}
\end{equation}

\noindent
Inserting the right hand side of Eq. (\ref{eq:steady}) in Eq. (\ref{eq:Gxout})
and carrying out the summations, one finds that
$G^{\rm out}(x)$ satisfies

\begin{equation}
G^{\rm out}(x) = x G^{\rm out}(1-p+px).
\end{equation}

\noindent
Differentiating both sides with respect to $x$, we obtain

\begin{equation}
\frac{d}{dx} G^{\rm out}(x) = G^{\rm out}(1-p+px) + x p \frac{d}{dx} G^{\rm out}(1-p+px).
\end{equation}

\noindent
Taking the $n^{\rm th}$ derivative we obtain

\begin{eqnarray}
\frac{d^n}{dx^n} G^{\rm out}(x) 
&=&
n p^{n-1} \frac{d^{n-1}}{dx^{n-1}} G^{\rm out}(1-p+px)
\nonumber \\
&+& x p^n \frac{d^{n}}{dx^{n}} G^{\rm out}(1-p+px).
\end{eqnarray}

\noindent
Inserting $x=1$ enables us to express the $n^{\rm th}$ derivative of
$G^{\rm out}(x) \vert_{x=1}$
in terms of the $(n-1)^{\rm th}$ derivative, in the form

\begin{equation}
\frac{d^n}{dx^n} G^{\rm out}(x) \bigg\vert_{x=1}
= \frac{n p^{n-1}}{1-p^n}
\frac{d^{n-1}}{dx^{n-1}} G^{\rm out}(x) \bigg\vert_{x=1}.
\end{equation}

\noindent
Using this relation recursively for $n'=1,2,\dots,n$,
we obtain

\begin{equation}
\frac{d^n}{dx^n} G^{\rm out}(x) \bigg\vert_{x=1}
=
\frac{ n! p^{n(n-1)/2} }{\prod\limits_{r=1}^n (1-p^r)}.
\label{eq:Gn}
\end{equation}

\noindent
The generating function can be expressed in terms of a series expansion
around $x=1$, which takes the form

\begin{equation}
G^{\rm out}(x) = 
\sum_{n=0}^{\infty} \frac{(x-1)^n}{n!} \frac{d^n}{dx^n} G^{\rm out}(x) \bigg\vert_{x=1}.
\label{eq:Gserexp}
\end{equation}

\noindent
Inserting the right hand side of Eq. (\ref{eq:Gn}) into Eq. (\ref{eq:Gserexp}),
the generating function can be expressed in the form

\begin{equation}
G^{\rm out}(x) = 
1 + \sum_{n=1}^{\infty}
\frac{p^{n(n-1)/2}}{\prod\limits_{r=1}^{n} (1-p^r)}
(x-1)^n.
\label{eq:Goutx}
\end{equation}

\noindent
Using the relation

\begin{equation}
P(K_{\rm out}=k) = \frac{1}{k!} \frac{d^k}{dx^k} G^{\rm out}(x) \bigg\vert_{x=0},
\end{equation}

\noindent
we treat separately the cases of $k=0$ and $k \ge 1$.
For $k=0$ we obtain

\begin{equation}
P(K_{\rm out}=0) = 
1 +
\sum\limits_{n=1}^{\infty}
\frac{
(-1)^{n} 
p^{n(n-1)/2} }{\prod\limits_{r=1}^n (1-p^r)}.
\label{eq:Pko1k0}
\end{equation}

\noindent
Using Eqs. (\ref{eq:q-Psp}) and (\ref{eq:q-Pinf1}) we find that

\begin{equation}
P(K_{\rm out}=0) = (1,p)_{\infty}=0.
\end{equation}

\noindent
For $k \ge 1$ we obtain

\begin{equation}
P(K_{\rm out}=k) = 
\sum\limits_{n=k}^{\infty}
{ {n} \choose {k} } 
\frac{
(-1)^{n-k} 
p^{n(n-1)/2} }{\prod\limits_{r=1}^n (1-p^r)},
\label{eq:Pko1}
\end{equation}

\noindent
This is a closed form expression for the out-degree distribution.

In the limit of $p \rightarrow 1^{-}$ Eq. (\ref{eq:dPdtKout})
is reduced to the form

\begin{equation}
\frac{\partial}{\partial t} P_t(K_{\rm out}=k) =
- \frac{  P_t(K_{\rm out}=k) - P_t(K_{\rm out}=k-1)}{t+s+1}.
\label{eq:dPdtKout2}
\end{equation}

\noindent
The structure of Eq. (\ref{eq:dPdtKout2}) is identical to that of 
Eq. (\ref{eq:Nt2}),
which describes the time dependence of $P_t(N_{\rm down}=n)$.
This similarity reflects the fact that in the limit of $p \rightarrow 1^{-}$
each node is connected by outgoing links to all its downstream nodes.
Therefore, in this limit 
$P_t(K_{\rm out}=k) = P_t(N_{\rm down}=k)$,
namely

\begin{equation}
P_t(K_{\rm out}=k) =
\frac{1}{t_s} \sum_{m=0}^{k} \frac{ P_0(K_{\rm out}=m)
\left( \ln t_s \right)^{k-m} }{(k-m)!}.
\end{equation}

\noindent
The $n^{\rm th}$ factorial moment of $P(K_{\rm out}=k)$ can be expressed by

\begin{equation}
\langle K_{\rm out}(K_{\rm out}-1)\dots(K_{\rm out}-n+1) \rangle 
= \frac{\partial^k}{\partial x^k} G^{\rm out}(x) \bigg\vert_{x=1}.
\label{eq:fout}
\end{equation}

\noindent
In particular, the mean out-degree,
obtained from Eq. (\ref{eq:fout}) with $n=1$,
is given by

\begin{equation}
\langle K_{\rm out} \rangle = \frac{1}{1-p}.
\end{equation}

\noindent
The second factorial moment, obtained from Eq. (\ref{eq:fout})
with $n=2$, is given by

\begin{equation}
\langle K_{\rm out} (K_{\rm out}-1) \rangle = \frac{2p}{(1-p)^2(1+p)}.
\end{equation}

The variance of $P(K_{\rm out}=k)$ is thus given by

\begin{equation}
{\rm Var}(K_{\rm out}) =
\frac{p}{1-p}.
\label{eq:VarKout}
\end{equation}

\noindent
The $n^{\rm th}$ moment $\langle K_{\rm out}^n \rangle$ 
can be expressed in terms of the factorial moments,
in the form

\begin{equation}
\langle K_{\rm out}^n \rangle = 
\sum_{k=0}^n {n \brace k}
\langle K_{\rm out}(K_{\rm out}-1)\dots(K_{\rm out}-k+1) \rangle,
\end{equation}

\noindent
where $n \brace k$ is the Stirling number of the second kind.

In Appendix B we consider the behavior of the out-degree distribution in the
limit of $p \ll 1$. We show that in this limit, the distribution
$P(K_{\rm out}=k)$ converges towards a Poisson-like
distribution of the form

\begin{equation}
P(K_{\rm out}=k) =   \frac{ \left( \frac{p}{1-p} \right)^{k-1} }{ (k-1)! } 
e^{- \left(  \frac{p}{1-p} \right) }.
\end{equation}

\noindent
In Appendix C we consider the behavior of the out-degree distribution in the
limit of $p \simeq 1$. We show that in this limit the distribution
$P(K_{\rm out}=k)$ converges towards a Gaussian distribution.

\begin{figure}
\centerline{
\includegraphics[width=7.0cm]{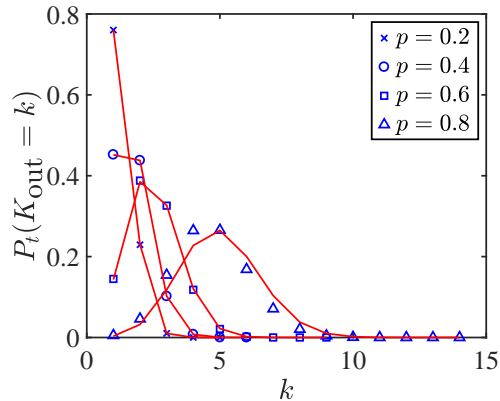}
}
\caption{
Analytical results (solid lines) for the
out-degree distribution
$P_t(K_{\rm out}=k)$,
of the corded DND network
as a function of $k$,
for $p=0.2$, $0.4$, $0.6$ and $0.8$.
The analytical results are found to be in good agreement with the results obtained
from computer simulations (symbols).
As $p$ is increased, the peak of the out-degree distribution shifts to 
the right and becomes broader.
}
\label{fig:4}
\end{figure}

In Fig. 
\ref{fig:4}
we present analytical results (solid lines) for the
out-degree distribution
$P(K_{\rm out}=k)$, 
of the corded DND network,
for $t=10^6$
and $p=0.2$, $0.4$, $0.6$ and $0.8$,
obtained from Eq. (\ref{eq:Pko1}).
The analytical results
are in very good agreement
with the results obtained from computer 
simulations (symbols).
It is found that as $p$ is increased, the 
out-degree distribution becomes broader
and nodes with higher out-degrees become more
probable.

\section{Discussion}

The corded DND model differs from its undirected counterpart is several ways.
Unlike the undirected model in which D may form probabilistic links to all the neighbors
of M, in the directed model it may form (directed) probabilistic links only to the outgoing
neighbors of M and not to the incoming neighbors of M.
In the undirected network each pair of nodes is connected by at least
one path. 
The distribution of shortest path lengths (DSPL) in the undirected ND
network was recently studied
\cite{Steinbock2017}.
It was found that in the long time limit the mean distance $\langle L \rangle_t$ 
scales like $\langle L \rangle_t \sim \ln N_t$, where $N_t$ is the network size at time $t$.
The mean distance thus scales logarithmically with the network size, which means that the
corded ND network is a small world network.
In contrast, in the corded DND network, 
from each node one can access 
along directed paths
only those older nodes that reside along the same branch of the backbone
tree. As a result, the probability that two random nodes are connected
by a directed path scales like $\ln N_t / N_t$.
Therefore, the corded DND network is not a small world network.

In directed networks it is important to distinguish between the
in-degree distribution and the out-degree distribution. We find that the
in-degree distribution follows a shifted power-law distribution.
The out-degree distribution 
converges to a Poisson distribution for $p \ll 1$
and to a Gaussian distribution for $p \simeq 1$.
Unlike the undirected ND network,
the DND network does
not exhibit a structural phase transition from a sparse phase to a dense phase.
Its structure over the whole range of
$0 < p < 1$ 
resembles the sparse phase of the undirected network and it does not exhibit 
a dense phase.

The corded DND model may be useful in the study of scientific
citation networks. In these networks each new paper emanates
from one or more papers which were previously published in
the literature. The earlier papers, which are cited in the new paper,
are analogous to the mother node in the corded DND network.
More specifically,
each paper is represented by a node and each citation
is represented by a directed link from the citing paper to the cited paper.
The in-degree of each node is the number of citations received by the
corresponding paper, while the out-degree is the number of papers 
that appear in the reference list at the end of the paper.
Clearly, the out-degree of a paper is easily accessible and is 
fixed once the paper is published.
In contrast, the in-degree of a paper is initially 
zero and it may grow as the paper gets
cited by subsequent papers.
The citations of each paper are spread throughout the scientific literature.
Gathering this information requires an effort.
It can be obtained from search engines such as the 
Institute of Scientific Information (ISI) Web of Knowledge
and Google Scholar.

Using large data sets of papers and citations, obtained from the ISI
and from the American Physical Society,
it was found that the distribution of in-degrees in scientific citation networks
is a fat-tail distribution that can be well 
fitted by a shifted power-law distribution
\cite{Redner1998}. 
The broad distribution of the number of citations received by different papers
can be attributed to the large variability in 
the quality and relevance of the published 
papers. However, it is also enhanced by the 
cumulative advantage mechanism 
\cite{Price1965,Price1976}, 
namely the fact that a paper that already 
received citations is more likely to 
receive additional citations in the future.
This correlation was studied in Ref. 
\cite{Redner2005} 
using all the papers published in the Physical Review journals between
1893 and 2003.
It was found that the expected number of 
citations a paper will receive during
year $t+1$ is linearly proportional 
to the number of citations it had up to year $t$.
This can be attributed to the fact that 
any existing citation provides an additional
channel through which the cited paper can be reached.
Unlike the in-degree distribution, it was found 
that the distribution of out-degrees 
in scientific citation networks is a narrow distribution. 
This can be explained by a few effects. 
First, the authors need a minimal number 
of citations in order to put their paper
in proper perspective, but excessive reference 
lists are counter-productive. 
A similar yet slightly different perspective, 
suggested by 
\cite{Golosovsky2017a},
is that authors try to comply with what is 
accepted in their research field in terms of references, 
implying that there is a feedback mechanism 
that forces the authors to adhere to some typical length of the 
reference list. This results in a relatively narrow 
distribution of out-degrees.
This is unlike the in-degrees that are 
determined by the citation dynamics of a paper, 
where the decision on whether to cite it comes 
from many uncoordinated authors and during 
a long period of time.

The corded DND model captures some essential properties 
of scientific citation networks.
It is a directed network whose links point 
from the citing paper to the cited paper.
The probabilistic links from the daughter node 
to outgoing neighbors of the mother node
correspond to the fact that a citation 
of a paper is often accompanied by citations to some
of the earlier papers that appear in its reference list. 
These probabilistic links also invoke
the preferential attachment mechanism, 
because the probability of a node to receive
such link is proportional to its in-degree. 
This is the mechanism that gives rise to the
power-law tail of the in-degree distribution.
The shift in the power-law degree distribution 
is due to the fact that the deterministic
links are formed by random attachment 
with no preference to high degree nodes.

The corded DND network provides insight on the structure
of the scientific citation networks. 
In particular, it indicates
that for a given paper, the typical number of papers that are connected
to it by directed paths of citations 
(in the past or future)
scales like $\ln N$, where
$N$ is the network size.
This implies that the scientific literature is highly fragmented
in the sense that most pairs of papers are not connected via 
chains of citations and thus the network is not a small-world network.
The distribution of the number of downstream nodes
sheds new light on the way the impact of a paper
may be evaluated, namely not only in terms of the direct
citations but also in terms of the cumulative effect of
all the subsequent indirect citations.

There are some features of citation networks that are not captured by the
corded DND model. In this model, at time $t$ all nodes are equally likely to
be selected as a mother node. However, in citation networks there is a 
phenomenon referred to as 'aging' of papers, namely an old paper is less
likely to be cited than a new paper \cite{Golosovsky2012,Golosovsky2017a}.
This can be incorporated into the model
by assigning an aging factor $a(\tau)$, where $\tau=t-t'$, 
which represents the likelihood that
a node that was formed at time $t'$ will be selected as a mother node at time $t$.
The function $a(\tau)$ is defined for $\tau \ge 1$ and is a monotonically
decreasing function of $\tau$.

An important difference between the model and actual citation networks is in
the way the time is defined. 
In empirical studies of citation networks, all the papers from a given year
are often grouped together. In some cases the internal order in the publication dates
within a given year is ignored. 
Since the number of papers published each year increases exponentially
over the last seven decades, this approach makes it difficult to compare
the properties of the network in different periods.
In the corded DND model, the clock is advanced by one
time unit each time a new paper appears.
This is a convenient choice because papers (or nodes) appear in discrete
units and thus it is most suitable to use discrete time.
This choice of time is also invariant to the actual rate of publication of
papers, which can be one paper per hour, day, week or month depending
on the size of the scientific field and the number of journals included in
the analysis.
We believe that using an internal clock based on a simple counting of the
papers under study in the order of their dates of publication
is likely to simplify the analysis.
Moreover, it is likely to be insensitive to the
increasing rate of publication and provide a more consistent picture over long
periods of time.
Alternatively, one can use the model in a way that accounts for the variable yearly
production of papers, by labeling the nodes accordingly.

In the context of citation networks the mother node can be considered as the
primary reference of the new paper. This can be the reference whose reading
by the author of the new paper inspired its writing. In citation networks there
may be several primary references in the reference list of a given paper, 
sometimes referred as direct (as opposed to secondary or indirect) citations 
\cite{Golosovsky2017a}.
It would be interesting to generalize the corded DND model to account for 
this possibility. In such a generalized model there will be deterministic
links to several random nodes and probabilistic links to their outgoing neighbors.
Such a model may not be analytically tractable, 
but the exact results of the current single-primary-reference model 
could be used as a starting point for a systematic approximation scheme, 
to be verified by extensive computer simulations. 
The key insight is that if indeed many such primary 
references motivate a certain research they are rather 
far away from each other in the network, or otherwise 
they could be considered part of the same sub-cluster 
with one of them being a single primary source. 

In the corded DND model it is essentially assumed that all nodes
are born equal and the accumulation of incoming links is a 
random process. Of course, as scientists who write papers,
we do not actually believe that this is indeed the case.
This aspect is sometimes referred to as initial attractivity 
\cite{Price1976,Golosovsky2018}
or fitness distributions of papers 
\cite{Golosovsky2017b}.
It would be interesting to study a modified version of the model,
in which each node, upon formation, is endowed with a
'citability index' $0 < \sigma < 1$, which would account for the
likelihood that this node will be selected for duplication at a later
time. The distribution of citability indices would represent the
variability in the quality, interest and relevance of the published papers.

\section{Summary}

We have studied a theoretical model of a directed network,
which grows by a node duplication mechanism.
In this model, 
referred to as the corded DND model,
at each time step a random mother node is 
duplicated.
The daughter node acquires a directed link to the mother node,
and  with probability $p$ a link to each
one of its outgoing neighbors.
We obtained exact analytical results for the time dependent distributions
of the in-degrees and the out-degrees of nodes in the resulting network.
The out-degrees exhibit a narrow distribution, 
which  converges to a Poisson distribution 
in the limit of $p \ll 1$
and to a Gaussian distribution in the limit
of $p \simeq 1$.
The in-degrees follow a shifted power-law distribution,
which means that the network is asymptotically a scale-free network,
with a power-law degree distribution of the form
$P(K_{\rm in}=k) \sim k^{-\gamma}$
where $\gamma = 1+1/p$.
The mean degree is identical for the in-degree and out-degree distributions
and in the asymptotic it converges to $\langle K \rangle = 1/(1-p)$
for the whole range of $0 < p < 1$.
This is in contrast to the undirected corded ND network, which 
exhibits a structural phase transition at $p=1/2$ 
between the sparse network regime of $p < 1/2$
and the dense network regime of $p > 1/2$,
in which the mean degree diverges in the large network limit
\cite{Lambiotte2016,Bhat2016}.

Since the network is directed not all pairs of nodes are connected by
directed paths even though  the corresponding undirected network
consists of a single connected component.
To analyze the connectivity of the network we
derived master equations for the distribution
of the number of upstream nodes
$P_t(N_{\rm up}=n)$,
and for the distribution of the number of downstream nodes
$P_t(N_{\rm down}=n)$.
The two distributions are very different from each other.
In the asymptotic limit, $P_t(N_{\rm up}=n) \sim 1/n^2$,
while $P_t(N_{\rm down}=n)$ is given by a sum of Poisson-like terms.
However, by a conservation law the two distributions satisfy
$\langle N_{\rm up} \rangle = \langle N_{\rm down} \rangle \sim \ln N_t$.
This means that in the large network limit 
the probability 
that a random pair of nodes
$i$ and $j$ are connected by a directed path from $i$ to $j$
scales like
$\ln N_t/N_t$.
Thus, the fraction of pairs of nodes that are connected by directed
paths diminishes like $\ln N_t/N_t$ as the network size increases,
while most pairs of nodes are not connected by directed paths.
Therefore, the corded DND network is not a small-world network,
unlike the corded undirected node duplication network.

\vspace{0.1in}

\noindent
This work was supported by the Israel Science Foundation
grant no. 1682/18.

\appendix
\setcounter{section}{0}

\section{A useful identity based on the q-Pochhammer symbol}

In this Appendix we prove a mathematical identity that will be useful
for the analysis of the out-degree distribution in the limit of $p \ll 1$.
To this end, we use the
q-Pochhammer symbol, 
which is defined by
\cite{Koepf1998}

\begin{equation}
(\xi;p)_n = \prod_{k=0}^{n-1} (1 - p^k \xi).
\label{eq:q-P}
\end{equation}

\noindent
For $n=\infty$ it can be expressed by

\begin{equation}
(\xi;p)_{\infty} = \prod_{k=0}^{\infty} (1 - p^k \xi),
\label{eq:q-Pinf0}
\end{equation}

\noindent
or by

\begin{equation}
(\xi;p)_{\infty} = (1-\xi) \prod_{k=1}^{\infty} (1 - p^k \xi).
\label{eq:q-Pinf1}
\end{equation}

\noindent
Using the q-binomial theorem, one can represent the q-Pochhammer
symbol as a sum of the form
\cite{Olver2010}

\begin{equation}
(\xi;p)_n = \sum_{k=0}^{n} 
{n \brack k}_p
p^{ k(k-1)/2 } (-\xi)^k,
\label{eq:q-Pexp}
\end{equation}

\noindent
where 
${n \brack k}_p$
is the q-binomial coefficient,
given by

\begin{equation}
{n \brack k}_p
=
\frac{ (p;p)_n }{ (p;p)_k (p;p)_{n-k} }.
\end{equation}

\noindent
In the limit of $n \rightarrow \infty$

\begin{equation}
{n \brack k}_p
\rightarrow
\frac{ 1 }{ (p;p)_k }.
\end{equation}

\noindent
Inserting this result in Eq. (\ref{eq:q-Pexp}), 
we obtain

\begin{equation}
(\xi,p)_{\infty} = \sum_{k=0}^{\infty} \frac{ (-1)^k p^{k(k-1)/2} }{ (p;p)_k } \xi^k.
\label{eq:q-Psum}
\end{equation}

\noindent
The term $(p;p)_k$ can be written explicitly in the form

\begin{equation}
(p;p)_k = \prod_{r=1}^k (1-p^r).
\label{eq:ppk}
\end{equation}

\noindent
Note that for $k=0$, $(p;p)_0=1$, in analogy to the fact that $0!=1$.
Inserting the expression for $(p;p)_k$ from Eq. (\ref{eq:ppk})
into Eq. (\ref{eq:q-Psum}), we obtain

\begin{equation}
(\xi,p)_{\infty} = 
1 +
\sum_{k=1}^{\infty} 
\frac{ (-1)^k p^{k(k-1)/2} }{ \prod_{r=1}^k (1-p^r) } \xi^k.
\label{eq:q-Psp}
\end{equation}

\noindent
Splitting the right hand side of Eq. (\ref{eq:q-Psp}) into two summations,
we obtain

\begin{equation}
(\xi,p)_{\infty} = 
1 +
\sum_{k=1}^{\infty} 
\frac{ (-1)^k p^{k(k+1)/2} }{ \prod_{r=1}^k (1-p^r) } \xi^k
-
\sum_{k=1}^{\infty} 
\frac{ (-1)^{k-1} p^{k(k-1)/2} }{ \prod_{r=1}^{k-1} (1-p^r) } \xi^k.
\end{equation}

\noindent
Re-adjusting the lower bound of the second sum to $k=0$
and combining the summations we obtain

\begin{equation}
(\xi,p)_{\infty} = 
(1-\xi)
\left[ 1 +
\sum_{k=1}^{\infty} 
\frac{ (-1)^k p^{k(k+1)/2} }{ \prod_{r=1}^k (1-p^r) } \xi^k \right].
\label{eq:q-Pfinal}
\end{equation}

\noindent
Comparing the right hand sides of
Eqs. (\ref{eq:q-Pinf1}) and (\ref{eq:q-Pfinal}), we find that

\begin{equation}
\prod_{k=1}^{\infty} (1-p^k \xi) = 
1 +
\sum_{k=1}^{\infty} 
\frac{ (-1)^k p^{k(k+1)/2} }{ \prod_{r=1}^k (1-p^r) } \xi^k.
\label{eq:unnamed}
\end{equation}

\section{The out-degree distribution in the limit of $p \ll 1$}

The generating function $G^{\rm out}(x)$ of the out-degree distribution,
given by Eq. (\ref{eq:Goutx}), can be 
expressed in the form

\begin{equation}
G^{\rm out}(x) = 
1 +
\sum_{k=1}^{\infty} 
\frac{ (-1)^k p^{k(k+1)/2} }{ \prod\limits_{r=1}^k  (1-p^r) } 
\left( \frac{1-x}{p} \right)^k.
\end{equation}

\noindent
Using Eq. (\ref{eq:unnamed}) we obtain

\begin{equation}
G^{\rm out}(x) = \prod_{k=0}^{\infty}
\left[ 1 - p^k ( 1-x ) \right].
\end{equation}

\noindent
From this expression, one can obtain the moment generating function
$M(x)$, which takes the form

\begin{equation}
M(x) = G^{\rm out}(e^{x}) =\prod_{k=0}^{\infty}
\left[ 1 + p^k (e^x - 1) \right]
\end{equation}

\noindent
and the cumulant generating function, 
$H(x) = \ln M(x)$,
which takes the form

\begin{equation}
H(x) = 
\sum_{k=0}^{\infty}
\ln \left[ 1 + p^k (e^x - 1) \right].
\end{equation}

\noindent
Expressing the logarithmic function 
as a power series in the small parameter
$p^k(e^{x}-1)$, we obtain

\begin{equation}
H(x) = 
\sum_{k=0}^{\infty}
\sum_{n=1}^{\infty}
\frac{(-1)^{n+1}}{n}
\left[ p^k (e^x - 1) \right]^n.
\end{equation}

\noindent
Writing out the binomial expansion, we obtain

\begin{equation}
H(x) = 
-
\sum_{k=0}^{\infty}
\sum_{n=1}^{\infty}
\frac{p^{kn}}{n}
\sum_{j=0}^{n}
{ {n} \choose {j} } (-1)^j e^{jx}.
\end{equation}

\noindent
Summing up over $k$, the expression is simplified to 

\begin{equation}
H(x) = 
-
\sum_{n=1}^{\infty}
\frac{1}{n(1-p^n)}
\sum_{j=0}^{n}
{ {n} \choose {j} } (-1)^j e^{jx}.
\end{equation}

\noindent
The $r^{\rm th}$ cumulant, $C_r$, of 
the out-degree distribution, $P(K_{\rm out}=k)$
is given by

\begin{equation}
C_r(K_{\rm out}) = \frac{\partial^r}{\partial x^r} H(x) \bigg\vert_{x=0}.
\end{equation}

\noindent
Carrying out the differentiation, we obtain

\begin{equation}
C_r(K_{\rm out}) =
-
\sum_{n=1}^{\infty}
\frac{1}{n(1-p^n)}
\sum_{j=0}^n 
{ {n} \choose {j} } (-1)^j j^r.
\label{eq:kappar2}
\end{equation}

\noindent
Expressing the right hand side of Eq.
(\ref{eq:kappar2}) in terms of the Stirling
numbers of the second kind, we obtain

\begin{equation}
C_r(K_{\rm out}) =
\sum_{n=1}^r 
{r \brace n}
\frac{(-1)^{n+1} (n-1)!}{1-p^n}. 
\label{eq:kappar3}
\end{equation}

\noindent
For $r=1$ we obtain

\begin{equation}
C_1(K_{\rm out}) = \frac{1}{1-p}.
\end{equation}

For $r \ge 2$ we expand the right hand side of 
Eq. (\ref{eq:kappar3}) in powers of $p$, and obtain

\begin{equation}
C_r(K_{\rm out}) =
\frac{1}{1-p}
- \sum_{n=2}^r 
{r \brace n}
(-1)^n (n-1)!  + O(p^2).
\label{eq:kappar4}
\end{equation}

\noindent
Using the identity

\begin{equation}
\sum_{n=1}^r 
{r \brace n}
(-1)^n (n-1)!  = 0,
\label{eq:kappar5}
\end{equation}

\noindent
we find that for $r \ge 2$

\begin{equation}
C_r(K_{\rm out}) =
\frac{p}{1-p}
+ O(p^2),
\label{eq:kappar4p}
\end{equation}

\noindent
regardless of the order $r$.
It is thus found that 
for $p \ll 1$ the equality
$C_1(K_{\rm out}) = C_r(K_{\rm out})$
is valid for all values of $r \ge 2$.
However, the equality occurs for the shifted random variable
$K_{\rm out}-1$.
This implies that $K_{\rm out}-1$ follows a Poisson distribution
whose mean is $p/(1-p)$. 
Therefore, in the limit of $p \ll 1$ the out degree distribution,
$P(K_{\rm out}=k)$, follows a shifted Poisson distribution
with mean degree of $1/(1-p)$.

\section{The out-degree distribution in the regime of $p \simeq 1$}

In the analysis of the out-degree distribution,
$P(K_{\rm out}=k)$, in the
dense network limit, it is convenient to
use the parameter $q=1-p$.
Using Eq. (\ref{eq:kappar3}) it is found that
the first two cumulants 
of $P(K_{\rm out}=k)$
are given by

\begin{equation}
C_1(K_{\rm out}) = \frac{1}{q}
\end{equation}

\noindent
and

\begin{equation}
C_2(K_{\rm out}) = \frac{1}{q} \left( \frac{1-q}{2-q} \right),
\end{equation}

\noindent
while higher order cumulants are given by

\begin{equation}
C_r(K_{\rm out}) =
\frac{1}{q}
\left[ 1 + \sum_{n=1}^{r-1} 
{r \brace {n+1}}
\frac{(-1)^n n!}{1+\sum\limits_{j=1}^n (1-q)^j} 
 \right].
\label{eq:kappar6}
\end{equation}

\noindent
Expanding the right hand side of Eq. (\ref{eq:kappar6}) in powers
of $q$ we obtain that for $r \ge 3$

\begin{equation}
C_r(K_{\rm out}) =
\frac{1}{q}
\left[ B_{r-1} - \frac{1}{2} B_{r-1} q + O(q^2) \right],
\label{eq:kappar7}
\end{equation}

\noindent
where $B_r$ are the Bernoulli numbers, given by

\begin{equation}
B_r = \sum_{n=0}^r 
{r \brace n}
\frac{(-1)^n n!}{n+1} .
\end{equation}

\noindent
The mean of the out-degree distribution, $P(K_{\rm out}=k)$,
in the long time limit, is given by

\begin{equation}
\langle K_{\rm out} \rangle = C_1(K_{\rm out}) = \frac{1}{q},
\label{eq:KmeanG}
\end{equation}

\noindent
while the variance is

\begin{equation}
{\rm Var}(K_{\rm out}) 
= 
C_2(K_{\rm out}) = \frac{1}{q} \left( \frac{1-q}{2-q} \right).
\label{eq:KvarG}
\end{equation}

\noindent
Therefore, the standardized random variable

\begin{equation}
X = \frac{K_{\rm out}- \langle K_{\rm out} \rangle}
{ \sqrt{ {\rm Var}(K_{\rm out}) }}
\end{equation}

\noindent
satisfies

\begin{equation}
C_1(X) = 0,
\end{equation}

\noindent
and

\begin{equation}
C_2(X) = 1,
\end{equation}

\noindent
while for $r \ge 3$

\begin{equation}
C_r(X) = 
\frac{ [q(2-q)]^{r/2} }{ q(1-q)^{r/2} } 
\left[ B_{r-1} - \frac{1}{2} B_{r-1} q + O(q^2) \right].
\label{eq:CrX}
\end{equation}

\noindent
Expanding 
the pre-factor of Eq. (\ref{eq:CrX})
in powers of $q$ we obtain

\begin{equation}
C_r(X) = 
2 B_{r-1} (2q)^{\frac{r}{2} - 1} + O\left( q^{\frac{r}{2}} \right).
\end{equation}

\noindent
Thus, in the limit of $q \rightarrow 0$, the cumulants of orders $r \ge 3$ vanish.
This means that in this limit the random variable $X$ follows the standard Gaussian
distribution, with $C_1(X)=0$, $C_2(X)=1$ and
$C_r(X)=0$ for $r \ge 3$.
Accordingly, in the limit of $p \simeq 1$,
the out-degree distribution 
$P(K_{\rm out}=k)$ 
converges towards a Gaussian distribution whose mean
$\langle K_{\rm out} \rangle$
is given by Eq. (\ref{eq:KmeanG})
and its variance ${\rm Var}(K_{\rm out})$
is given by Eq. (\ref{eq:KvarG}).


\clearpage
\newpage



\begin{thebibliography}{10}


\bibitem{Albert2002}
Albert R and   Barab\'asi A -L 2002
{\it Rev. Mod. Phys.} {\bf 74} 47  

\bibitem{Caldarelli2007}
Caldarelli G 2007
{\it Scale free networks: complex webs in nature and technology} 
(Oxford: Oxford University Press) 

\bibitem{Havlin2010}
Havlin S and Cohen R 2010
{\it Complex Networks: Structure, Robustness and Function}
(Cambridge: Cambridge University Press) 

\bibitem{Newman2010}
Newman M E J  2010
{\it Networks: an Introduction} 
(Oxford: Oxford University Press).

\bibitem{Estrada2011b}
Estrada E 2011
{\it The Structure of Complex Networks: Theory and Applications}
(Oxford: Oxford University Press).

\bibitem{Barrat2012}
Barrat A, Barth\'elemy M and Vespignani A 2012
{\it Dynamical Processes on Complex Networks}
(Cambridge: Cambridge University Press)



\bibitem{Barabasi1999}
Barab\'asi A -L and Albert R 1999
{\it Science} {\bf 286} 509  

\bibitem{Jeong2000}
Jeong H, Tombor B, Albert R, Oltvai Z N and Barab\'asi A -L 2000
{\it Nature} {\bf 407} 651  

\bibitem{Krapivsky2000}
Krapivsky P L, Redner S and Leyvraz F 2000
{\it Phys. Rev. Lett.} {\bf 85} 4629   

\bibitem{Krapivsky2001}
Krapivsky P L and Redner S 2001
{\it Phys. Rev. E} {\bf 63} 066123 

\bibitem{Vazquez2003}
V\'azquez A 2003
{\it Phys. Rev. E} {\bf 67} 056104 



\bibitem{Milgram1967}
Milgram S 1967
{\it Psychology Today} {\bf 1} 61 

\bibitem{Watts1998}
Watts D and Strogatz S 1998
{\it Nature} {\bf 393} 440  

\bibitem{Chung2002}
Chung F and Lu L 2002
{\it Proc. Nat. Acad. Sci. USA} {\bf 99} 15879  


\bibitem{Chung2003}
Chung F and Lu L 2003
{\it Internet Mathematics} {\bf 1} 91  


\bibitem{Cohen2003}
Cohen R and Havlin S 2003
{\it Phys. Rev. Lett.} {\bf 90} 058701 








\bibitem{Bhan2002}
Bhan A, Galas D J and Dewey T G 2002 
{\it Bioinformatics} {\bf 18} 1486  


\bibitem{Kim2002}
Kim J, Krapivsky P L, Kahng B and Redner S 2002
{\it Phys. Rev. E} {\bf 66} 055101 

\bibitem{Chung2003b}
Chung F, Lu L, Dewey T G and Galas D J 2003
{\it J. Comput. Biol.} {\bf 10} 677 

\bibitem{Krapivsky2005}
Krapivsky P L and Redner S 2005
{\it Phys. Rev. E} {\bf 71} 036118 

\bibitem{Ispolatov2005}
Ispolatov I, Krapivsky P L and Yuryev A 2005
{\it Phys. Rev. E} {\bf 71} 061911 

\bibitem{Ispolatov2005b}
Ispolatov I, Krapivsky P L, Mazo I and Yuryev A 2005
{\it New J. Phys.} {\bf 7} 145

\bibitem{Bebek2006}
Bebek G,  Berenbrink P, Cooper C, Friedetzky T, Nadeau J and Sahinalp S C 2006
{\it Theor. Comput. Sci.} {\bf 369} 239  

\bibitem{Li2013}
Li S, Choi K P and Wu T 2013
{\it Theor. Comput. Sci.} {\bf 476} 94  




\bibitem{Lambiotte2016}
Lambiotte R, Krapivsky P L, Bhat U and Redner S 2016
{\it Phys. Rev. Lett.} {\bf 117} 218301  


\bibitem{Bhat2016}
Bhat U, Krapivsky P L, Lambiotte R and Redner S 2016
{\it Phys. Rev. E.} {\bf 94} 062302



\bibitem{Steinbock2017}
Steinbock C, Biham O and Katzav E 2017
{\it Phys. Rev. E} {\bf 96}  032301 

\bibitem{Toivonen2009}
Toivonen R, Kovanen L, Kivel\"a M, Onnela J -P, Saram\"aki J and Kaski K 2009
{\it Social Networks} {\bf 31} 240  




\bibitem{Granovetter1973}
Granovetter M 1973
{\it American Journal of Sociology} {\bf 78} 1360  

\bibitem{Newman2001b}
Newman M E J 2001
{\it Proc. Natl. Acad. Sci. USA} {\bf 98} 404  


\bibitem{Milo2002}
Milo R, Shen-Orr S, Itzkovitz S, Kashtan N, Chklovskii D and Alon U 2002
{\it Science} {\bf 298} 824  

\bibitem{Alon2006}
Alon U 2006
{\it An Introduction to Systems Biology: Design Principles of Biological Circuits}
(Chapman and Hall/CRC).



\bibitem{Molloy1995}
Molloy M and Reed B 1995
{\it Random Struct. Algorithms} {\bf 6} 161 


\bibitem{Molloy1998}
Molloy M and Reed B 1998
{\it Combinatorics, Probability and Computing} {\bf 7} 295 

\bibitem{Newman2001}
Newman M E J, Strogatz S H and Watts D J 2001
{\it Phys. Rev. E} {\bf 64} 026118 



\bibitem{Ohno1970}
Ohno S 1970
{\it Evolution by Gene Duplication} (New York: Springer-Verlag).


\bibitem{Teichmann2004}
Teichmann S A and  Babu M M 2004
{\it Nature Genetics} {\bf 36} 492   



\bibitem{Redner1998}
Redner S 1998
{\it Eur. Phys. J. B} {\bf 4} 131  

\bibitem{Redner2005}
Redner S 2005
{\it Physics Today} {\bf 58}  49  

\bibitem{Radicchi2008}
Radicchi F, Fortunato S and Castellano C 2008
{\it Proc. Natl. Acad. Sci. USA} {\bf 105} 17268  




\bibitem{Golosovsky2012}
Golosovsky M and Solomon S 2012
{\it Phys. Rev. Lett.} {\bf 109} 098701 

\bibitem{Golosovsky2017a}
Golosovsky M and Solomon S 2017
{\it Phys. Rev. E} {\bf 95} 012324 


\bibitem{Golosovsky2017b}
Golosovsky M 2017
{\it Phys. Rev. E} {\bf 96} 032306  


\bibitem{Peterson2010}
Peterson G J, Press\'e S and Dill K A 2010
{\it Proc. Natl. Acad. Sci. USA} {\bf 107 } 16023  



\bibitem{Smythe1995}
Smythe R T and Mahmoud H 1995
{\it Theory Probab. Math. Statist.} {\bf 51} 1

\bibitem{Drmota1997}
Drmota M and Gittenberger B 1997
{\it Random Struct. Alg.} {\bf 10} 421 

\bibitem{Drmota2005}
Drmota M and Hwang H -K 2005
{\it Adv. Appl. Probab.} {\bf 37} 321 


\bibitem{Olver2010}
Olver F W J, Lozier D M, Boisvert R F and Clark C W 2010
{\it NIST Handbook of Mathematical Functions} 
(Cambridge: Cambridge University Press).





\bibitem{Price1965}
de Solla Price D J 1965
{\it Science} {\bf 149} 510 

\bibitem{Price1976}
de Solla Price D J 1976
{\it J. Am. Soc. Inf. Sci.} {\bf 27} 292 

\bibitem{Golosovsky2018}
Golosovsky M 2018
{\it Phys. Rev. E} {\bf 97} 062310 

\bibitem{Koepf1998}
Koepf W 1998
{\it Hypergeometric Summation: 
An Algorithmic Approach to Summation and Special Function Identities}
(Braunschweig: Vieweg and Teubner Verlag).

\end{thebibliography}
\end{document}